\documentclass[twocolumn, astrosymb]{aastex701}
\hypersetup{linkcolor=blue,citecolor=blue,filecolor=cyan,urlcolor=blue}

\submitjournal{ApJ}
\usepackage{amsmath,amstext}
\usepackage{amssymb}
\usepackage{hyperref}
\usepackage{graphicx}

\begin{document}

%\title{Constraining radial diffusivity in surface flux transport model using insights from the three dimensional dynamo model}
\title{Constraining the radial decay timescale of solar surface magnetic field through a comparative study of data-assimilative 2D surface flux transport and 3D dynamo models}

\author[orcid=0000-0001-7957-1727,gname=Soumyadeep, sname=Chatterjee]{Soumyadeep Chatterjee}
\affiliation{Department of Physics, Indian Institute of Technology Kanpur, Kanpur 208016, India.}
\email[show]{soumyade@iitk.ac.in}  

\author[orcid=0000-0001-5388-1233,gname=Gopal, sname=Hazra]{Gopal Hazra} 
\affiliation{Department of Physics, Indian Institute of Technology Kanpur, Kanpur 208016, India.}
\email[show]{hazra@iitk.ac.in}

\begin{abstract}
The polar magnetic field is the most reliable precursor for predicting the amplitude of the solar cycle, and the 2D surface flux transport (SFT) model is widely used to reconstruct its evolution. Traditional 2D SFT models can not capture the surface-interior coupling of the surface field, causing delays in polar field reversals. This deficiency is conventionally corrected by adding a decay term $-B_r/\tau$ with a poorly constrained radial decay timescale $\tau$. Here, we present a self-consistent estimate of $\tau$ through a comparative study of radial flux transport in the 2D SFT model and the 3D kinematic dynamo model, STABLE. By keeping the same transport parameters for both models and assuming surface-interior coupling is diffusive, the poloidal field evolution equation reduces to an eigenvalue problem, which yields a spectrum of $\tau$ that decrease with increasing angular modes $l$. To capture realistic surface-interior coupling, we further perform data-assimilated 2D SFT simulations with real magnetograms and compare those results with that of the data-assimilated 3D STABLE model to constrain $\tau$ and effective decay modes. With our  choice of transport parameters, a value of $\tau=~2~\text{yr}$ keeps the surface dynamics of the two models consistent, and this timescale corresponds to the angular mode $l=8$ from our self-consistent estimate. We also perform 2D SFT simulations with only large-scale active regions, as the source. We find that $\tau=7~\text{yr}$ accurately captures the radial decay of the dipole mode ($l=1$) and removes the secular drift in the polar fields.

\end{abstract}

 %--- \uat{Cosmology}{343} --- \uat{High Energy astrophysics}{739} --- \uat{Interstellar medium}{847} --- \uat{Stellar astronomy}{1583} --- \uat{Solar physics}{1476}}


\section{Introduction}\label{sec:introduction}

Solar magnetic activity shapes the structure of the solar corona and drives space weather disturbances that affect our near-Earth environment \citep[e.g.,][]{SchrijverSoPh2003,  Schwenn2006, Pulkkinen2007, MackayLRSP2012, TemmerLRSP2021, nandy2021b}. The state of this activity is primarily measured by the number of sunspots on the solar surface. Sunspots appear on the solar surface as a result of an oscillatory dynamo operating in the solar convection zone\citep{Chou11, Charbonneau2020}. As sunspots decay, their magnetic flux is transported poleward, building up the polar fields. The strength of this polar field serves as the most reliable precursor for predicting the amplitude of the subsequent solar cycle \citep{schatten1978, petrovay2020, JiangJASTP2023}. 
 
The Surface Flux Transport (SFT) model, originally formulated by \cite{LeightonApJ1964}, provides a phenomenological framework for obtaining a reasonable reconstruction of the Sun's polar field. In this model, magnetic flux from newly emerged sunspots is passively transported by large-scale flows, while being simultaneously dispersed by supergranular diffusion \citep{WangSci1989, SheeleyLRSP2005, Jiang2014SSRv}. The large-scale flows that consist of the poleward meridional flow and differential rotation \citep{Hathway_2003ESAS} play the most important role in the build-up of the polar field. The poleward meridional flow is essential for concentrating the trailing-polarity flux into polar caps and generating a realistic estimate of the polar field \citep{DeVoreSoPh1984, WangApJ1991}, whereas the surface differential rotation that varies across latitudes stretches the surface poloidal magnetic field more strongly near the equator, as the equatorial region rotates faster than the poles. The accuracy of the SFT model in reproducing the observed polar field depends sensitively on the choice of these transport parameters. Considerable effort has been devoted to calibrating and optimizing the parameters governing these flows and supergranular diffusion. \cite{BaumannAA2004} explored the effect of varying these parameters on the polar field and unsigned flux on the solar surface. Recently, \citet{Lemerle_2015ApJ} and \citet{WhitbreadAA2017} employed a $\chi^2$ based optimization method, PIKAIA, to simultaneously constrain the parameters against observed magnetograms. \cite{PetrovayAA2019} further mapped the full parameter space in terms of polar field characteristics. Overall, the results of these optimization studies show that multiple combinations of parameters of the large-scale flows and diffusivity can produce identical results, as the surface dynamics is controlled by their ratio - the magnetic Reynolds number \citep{YeatesSSRv2023}. 

Besides the transport parameters, numerical modeling of the source term also plays an important role in the SFT model. Newly emerged active regions are typically modeled as symmetric bipolar Gaussian patches with numerically assigned flux, tilt angle, and separation. The assigned flux, tilt angles, and separation between two polarities are drawn either from observed distributions or directly assimilated from magnetograms \citep{BaumannAA2004, JiangAA2011II, YeatesSoPh2015,VirtanenAA2017}. However, observed sunspots are not simple symmetric bipolar regions. \cite{Iijima2019} and \cite{WangAA2021} presented that morphological asymmetry between the compact leading polarity and the more diffuse trailing polarity has a significant influence on the polar field amplitude and the net dipole contribution. Incorporating this asymmetry, along with observationally derived tilt angles and historical sunspot records, \cite{PalApJ2026} reproduced polar field evolution that is substantially closer to observations. \citet{Yeates2020} showed that the simplified bipolar approximation overestimates the axial dipole moment when all parameters are kept unchanged, by incorporating sunspot directly from HMI magnetogram as Space-weather HMI Active Region Patches (SHARPs). Recently, data assimilation techniques have also been employed to incorporate observed magnetic field directly into SFT simulations as an alternative to active regions as the source \citep{Upton_2014ApJ, Hickmann_2015SoPh, Dash_2024ApJ}.

Beyond these developments in source term modeling, SFT models have been widely employed for multi-decadal and century-long reconstructions of the solar magnetic field \citep{JiangAA2011II, VirtanenAA2017, UptonGeoRL2018, PalApJ2026}, for predicting the amplitude of upcoming solar cycles \citep{Cameron2016, Jiang2018ApJ}, as photospheric boundary conditions for coronal and heliospheric models \citep{Schrijver_2003SoPh, YeatesSoPh2015}, and as surface constraints for flux transport dynamo models \citep{CameronAA2012, bhowmik2018, LuoAA2026}. However, the reliability of all of these applications depends on a less explored aspect of the model, which is the radial decay timescale. The phenomenological SFT equation treats the photospheric field as a purely surface quantity and does not account for its coupling to the deeper convection zone. This omission leads to an unrealistically long memory of the system, causing secular drifts in the polar field when cycles of varying amplitude are simulated and, in some cases, suppressing polar field reversals altogether \citep{Schrijver2002ApJ, BaumannAA2006}. Accounting for this radial coupling is therefore critical for the long-term fidelity of SFT simulations.

To address this, \cite{Schrijver2002ApJ} proposed adding a radial decay term -$B_r/\tau$ to the SFT model, where $B_r$ and $\tau$ are the surface radial magnetic field and the radial decay time scale respectively. \citet{BaumannAA2006} provided a physically motivated justification for this term by deriving $\tau$ as $\sim 5$ yr using eigenmode analysis of the diffusion operator in a spherical shell with volume diffusivity 100 km$^2$ s$^{-1}$. \citet{VirtanenAA2017} applied the SFT model of \cite{YeatesSoPh2015} and extended the model with this decay timescale for solar cycles 21-24. They reported that $\tau$ is indispensable for reproducing the polar field reversal of weak cycles, such as the reversal of weak Cycle 24 following the stronger Cycle 23. Independent evidence for its necessity was also provided by \cite{PetrovayAA2019}, who systematically mapped the parameter space consisting of transport parameters and $\tau$, and showed that without a significant radial time scale ($\tau >$  10 yr), the global dipole moment reverses too late across all parameter regimes, whereas $\tau = 5 \ \text{to} \ 10$ yr yields solutions consistent with observations. Despite this progress, the precise value of $\tau$ and its potential cycle dependence remain poorly constrained, with published values ranging from 4.5 to 32 yr depending on the optimization metric, the representation of source term, and the flow profile employed \citep{Lemerle_2015ApJ, WhitbreadAA2017,  Yeates_2023SSRv}. Alternative mechanisms, such as cycle-to-cycle variations in tilt angle \citep{Cameron_2010ApJ, GolubevaMNRAS2023, PalApJ2023} and time-dependent meridional flow \citep{Wang2002ApJ}, have also been proposed to mitigate the secular drift problem, but the observed variability of the polar field across cycles has not been satisfactorily explained by these approaches alone.

The root of the uncertainty in determining $\tau$ lies in the absence of a self-consistent link between the surface decay and the radial transport of flux in the interior of the convection zone. \citet{BaumannAA2006} estimated $\tau$ assuming a uniform volume diffusivity of $100$~km$^2$s$^{-1}$ throughout the convection zone and a different turbulent diffusivity on the surface of the Sun, and subsequent optimization studies constrained $\tau$ along with transport parameters without addressing its self consistent physical origin. We address this issue in the present work through a comparative study of two-dimensional (2D) SFT model and 3D dynamo model STABLE \citep{Miesch_Dikpati_2014, Hazra_2017, KM17, HM18, Chatterjee2026ApJ}. Since the horizontal transport on the surface of the three-dimensional (3D) kinematic flux transport dynamo model STABLE obeys the same physics as the 2D SFT, we isolate the radial diffusive coupling from the 3D induction equation and derive a coupled surface-interior eigenvalue equation for the convection zone, which subsequently provides the value of $\tau$ for a given radial profile of turbulent diffusivity. This establishes, for the first time, a self-consistent bridge between the 2D surface description and the 3D interior dynamics. 

The STABLE dynamo model solves the magnetic induction equation in a 3D rotating spherical shell encompassing the solar convection zone. In this model, the poleward transported flux is carried into the interior convection zone by the subducting meridional flow beyond $\sim 75^{\circ}$ latitude and diffuses radially through the turbulent diffusivity in the interior, which is approximated in the 2D SFT through $-B_r/\tau$ term. The surface dynamics of the 3D and 2D models are therefore expected to diverge poleward of $\sim 75^{\circ}$ latitude without an accurate incorporation of the radial decay term $-B_r/\tau$ in 2D SFT model. We therefore validate our self-consistent estimate of the eigenvalue equation of the poloidal field, as described earlier, through two complementary approaches that probe two different aspects of this surface-interior coupling. In the first approach, we assimilate the observed surface magnetic field distribution within $\pm75^{\circ}$ latitude and keep the unsigned flux same in both the 3D and 2D models. The comparison of the polar fields (poleward of $~75^{\circ}$) from both models then reveals a suitable value of $\tau$ and the dominant length scale in terms spherical harmonic $l$ up to which radial decay timescale is important for matching the surface dynamics of the 2D with the 3D model. In the second approach, we consider a traditional 2D SFT model with only active regions as source and verify whether our self-consistent estimation of $\tau$ can rectify the secular drift problem in the 2D SFT model.

%In this paper, we systematically constrain the radial decay timescale $\tau$ for the two-dimensional (2D) SFT model. 

The plan of the paper is as follows. In the next section, we present the analytical prescription of the self-consistent estimate of $\tau$ using a physically motivated depth-dependent radial diffusivity profile. In the subsequent Section~\ref{sec:dataassimilatedSFT}, we carry out the first approach by assimilating observed daily distributions of the magnetic field by the Helioseismic and Magnetic Imager (HMI) in the 2D SFT model, covering Cycles 24-25, and comparing the polar field estimates from this data assimilated 2D SFT with the polar field obtained from the 3D STABLE model of \cite{Chatterjee2026ApJ}. In Section~\ref{sec:test_w_SFT_sharp}, we present the second approach using the 2D standard SFT model driven by SHARP active regions and compare the polar fields with the observed polar field from the Wilcox observatory (WSO). Finally, we present our conclusions in Section~\ref{sec:conclusion}.

\section{Analytical prescription to determine radial decay timescale ($\tau$)} \label{sec:derivtau}

The standard 2D SFT model \citep{BaumannAA2006, Jiang_2023} solves the equation,
\begin{align}
            \frac{\partial B_r}{\partial t}=&-\Omega(\theta)\frac{\partial B_r}{\partial \phi}-\frac{1}{R_{\odot}\sin{\theta}}\frac{\partial}{\partial \theta}[v(\theta)B_r \sin{\theta}]& \notag \\ 
           &+\frac{\eta_H}{R_{\odot}^2}\left[\frac{1}{\sin{\theta}} \frac{\partial}{\partial \theta}\left( \sin{\theta} \frac{\partial B_r}{\partial \theta}\right) + \frac{1}{ \sin^2{\theta}}\frac{\partial^2 B_r}{\partial \phi^2} \right]& \notag \\ 
           &+S_{2D}(\theta,\phi,t)+D(\eta),
           \label{eq:2dsft}
\end{align}
where $B_r(\theta,\phi,t)$ is the radial component of the magnetic field on the surface of the Sun, expressed as a function of  heliographic colatitude ($\theta$) and longitude ($\phi$), respectively. $\Omega(\theta)$ is the differential rotation rate, $v(\theta)$ is the poleward meridional flow speed, $\eta_H$ is the horizontal supergranular diffusivity, and $S_{2D}(\theta,\phi,t)$ is the source term representing the emergence of new active regions or observed surface magnetograms. In this model, the diffusive coupling of the surface magnetic field with the deeper convection zone is not explicitly accounted for; instead, it is approximated by a linear decay term $D(\eta) = - B_r/\tau$, where $\tau$ is the decay timescale associated with the radial diffusivity, $\eta$. The omission of this decay term leads to an unrealistically long memory of the surface field, causing secular drifts in the polar field when cycles of varying amplitude are simulated \citep{Schrijver2002ApJ, BaumannAA2006}. In this section, we derive an analytical prescription for the time scale $\tau$ for a given radial profile of $\eta(r)$ by comparing the 2D SFT model with 3D kinematic mean field flux transport dynamo model STABLE \citep{Chatterjee2026ApJ}.  

The 3D kinematic flux transport dynamo model STABLE \citep{Miesch_Dikpati_2014, HCM17, HM18, Chatterjee2026ApJ} self-consistently couples the interior dynamo to the surface flux evolution through axisymmetric large-scale flows, turbulent diffusion, and the explicit emergence and dispersal of sunspots on the photosphere. In this model, the evolution of the magnetic field $\mathbf{B}(r,\theta, \phi,t)$ is governed by the induction equation: 

\begin{equation}
    \frac{\partial \mathbf{B}}{\partial t} = \nabla \times (\mathbf{V} \times \mathbf{B} - \eta \nabla \times \mathbf{B}) + \mathbf{S}_{3D},\label{eq:3dstable}
\end{equation}

where $\mathbf{V}$ comprises the 3D differential rotation and meridional circulation in the full spherical geometry of the solar convection zone, $\eta$ is the turbulent diffusivity, and $\mathbf{S}_{3D}$ is a parametrized source term governing the deposition of sunspots in response to the deep toroidal field, encapsulating the Babcock–Leighton mechanism. The full spherical geometry stretches from $0.69R_{\odot}$ (or $0.7R_{\odot}$) to $R_{\odot}$ in radius, $0^\circ$ to $180^\circ$ in latitude, and $0^\circ$ to $360^\circ$ in longitude, where $R_{\odot}$ is the solar radius. The magnetic field, $\mathbf{B}$ in the Equation~\ref{eq:3dstable} is solved in terms of toroidal ($A$) and poloidal ($C$) potentials using

\begin{equation} 
    \mathbf{B} = \nabla \times (A \hat{r}) + \nabla \times \nabla \times (C \hat{r}),\label{eq:ckdecompose}
\end{equation}
with a radial boundary conditions on the surface ($A = \partial C/\partial r = 0$), and perfectly conducting boundary conditions ($C = \partial A/\partial r = 0$) at the base of the convection zone, as described in \cite{Miesch_Teweldebirhan_2016}.

In the presence of the same meridional flow profile and the same surface diffusivity, and if we assume that the coupling between the surface and interior convection zone is dominated by diffusion, then Equation~\ref{eq:3dstable} reduces to
\begin{equation}
\frac{\partial \mathbf{B}}{\partial t} = -\nabla \times \big(\eta(r)\,\nabla \times \mathbf{B}\big),
\label{eq:diffusion}
\end{equation}
where $\eta$ is taken to be a function of radial coordinate r alone. Since the SFT model deals exclusively with $B_r$ at the solar surface, it is sufficient to consider the decay of a purely poloidal field. Accordingly, we retain only the second term in the Equation \ref{eq:ckdecompose} and the radial component takes the form
\begin{equation}
B_r = -\frac{L^2 C}{r^2},
\label{eq:Br_poloidal}
\end{equation}
where $L^2$ is the angular part of the Laplacian operator. Substituting Equation~(\ref{eq:Br_poloidal}) into the radial component of Equation~(\ref{eq:diffusion}) and simplifying, we obtain
\begin{equation}
\frac{\partial C}{\partial t} = \eta(r)\left(\frac{\partial^2 C}{\partial r^2} + \frac{L^2 C}{r^2}\right).
\label{eq:C_evolution}
\end{equation}
 
Following \citet{Elsasser1946PhRv, BaumannAA2006}, we decompose $C$ into orthogonal decay modes as 

\begin{equation}
C(r,\theta,\phi,t) = \sum_{n=0}^{\infty}\sum_{l=1}^{\infty}\sum_{m=-l}^{l} R_{nl}(r)\,Y_l^m(\theta,\phi)\,T_{ln}(t),
\label{eq:C_expansion}
\end{equation}
where $R_{nl}(r)$ are the radial eigenfunctions, $Y_l^m(\theta,\phi)$ are the spherical harmonics of degree $l$ and order $m$, and $T_{ln}(t)$ are the temporal decay functions. Substituting it into Equation~(\ref{eq:C_evolution}) and using $L^2 Y_l^m = -l(l+1)\,Y_l^m$, the temporal part separates as
\begin{equation}
T_{ln}(t) = \exp(-\lambda_{nl}\,t),
\label{eq:temporal_decay}
\end{equation}
where $\lambda_{nl}$ is the eigenvalue, and the corresponding decay timescale is $\tau_{nl} = \lambda_{nl}^{-1}$. Subsequently, the radial functions $R_{nl}(r)$ satisfy the ordinary differential equation
\begin{equation}
R_{nl}'' + \left[\frac{\lambda_{nl}}{\eta(r)} - \frac{l(l+1)}{r^2}\right]R_{nl} = 0,
\label{eq:radial_ODE}
\end{equation}
subject to the boundary conditions $R_{nl}'(R_\odot) = 0$ at the solar surface and $R_{nl}(R_{\rm b}) = 0$ at the base of the convection zone, where we take $R_{\rm b} = 0.69\,R_\odot$.
Thus, for a given $\eta(r)$, Equation~\ref{eq:radial_ODE} becomes an eigenvalue problem whose solutions yield the eigenvalues $\lambda_{nl}$ and correspondingly determine the decay time scale $\tau_{nl}$ for each radial and angular wavenumbers $n$ and $l$.  

\begin{figure}[!htbp]
    \centering
    \includegraphics[width=0.5\textwidth]{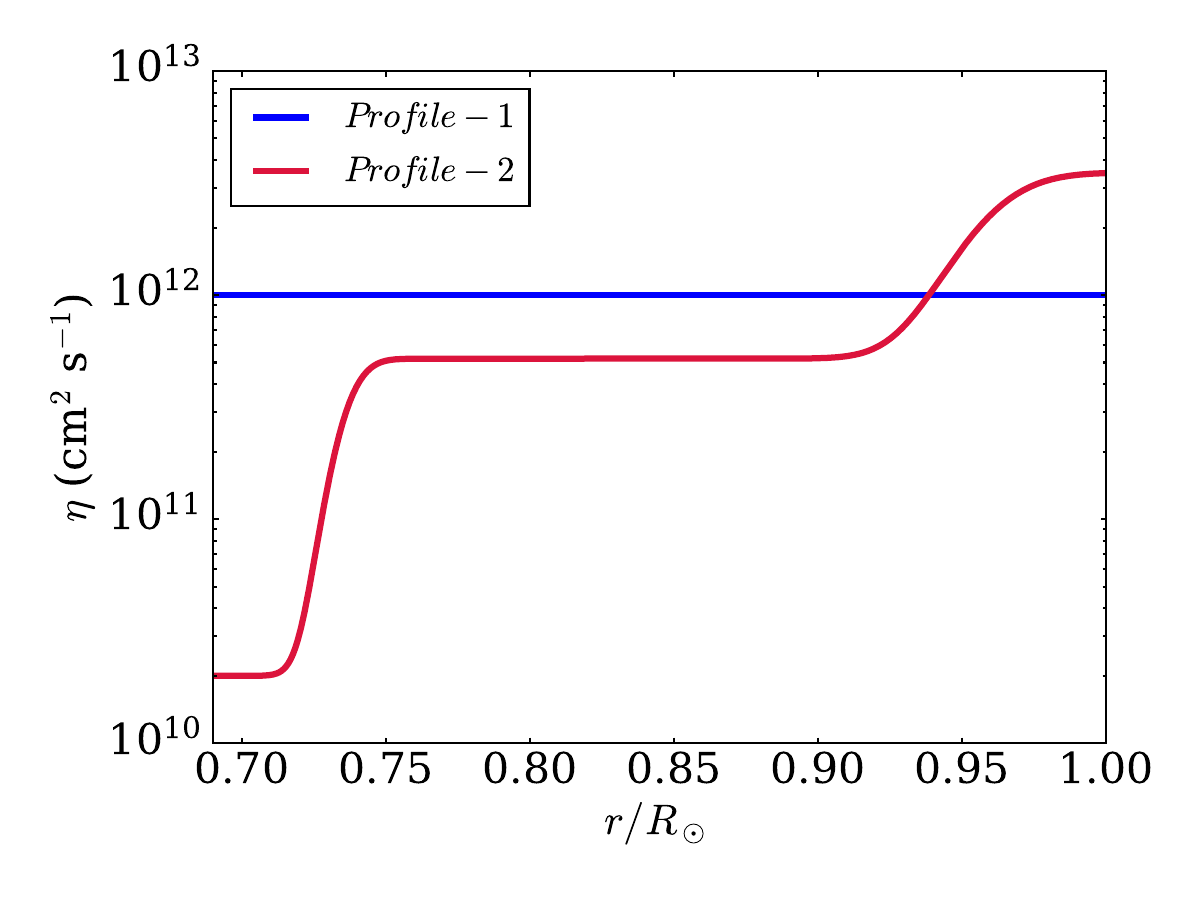}
    \caption{Profiles of radial diffusivity $\eta(r)$.}
    \label{fig:diffusivity_profiles}
\end{figure}

In this paper, we consider two profiles for $\eta(r)$, as shown in Figure~\ref{fig:diffusivity_profiles}. 
\textit{Profile}-1 adopts a constant value of $\eta=1\times10^{12} \text{cm}^2\text{s}^{-1}$ throughout the convection zone. However, \textit{Profile}-2 employs a two-step variation analytically expressed as 

\begin{align}
    \eta = \eta_c + \frac{\eta_{mid}}{2} \left[1 + \text{erf}\left(2\frac{r - r_{da}}{d_a}\right)\right] \notag \\
    + \frac{\eta_{H}}{2} \left[1 + \text{erf}\left(2\frac{r - r_{db}}{d_b}\right)\right], \label{eq:diffpro2}
\end{align}
where $\eta_c$, $\eta_{mid}$, and $\eta_{H}$ are the diffusivities at the bottom, middle, and top of the convection zone, respectively. We set $\eta_{H} = 3\times10^{12}\,\text{cm}^2\text{s}^{-1}$, $\eta_{mid} = 5\times10^{11}\,\text{cm}^2\text{s}^{-1}$, $\eta_c = 2\times10^{10}\,\text{cm}^2\text{s}^{-1}$, $r_{da} = 0.735R_\odot$, $r_{db} = 0.956R_\odot$, $d_a = 0.021R_\odot$, and $d_b = 0.05R_\odot$. The value of $\eta_{top}$ is consistent with the previous studies \citep{Jiang2014SSRv, Chatterjee2026ApJ}, and a high $\eta_{mid}$ ensures that diffusion dominates over advection by the meridional flow within the bulk of the convection zone \citep{yeates2008}. Therefore we solve the boundary value problem defined by Equation~\ref{eq:radial_ODE} numerically for each mode $l$ and $n$ using a shooting method combined with the Brent root-finding algorithm \citep{Brent1973} to determine the eigenvalues $\lambda_{nl}$.
 
Table~\ref{tab:decay_times} lists the values of $\tau_{nl}$ obtained by solving Equation~(\ref{eq:radial_ODE}) for both diffusivity profiles for $n = 0$ and $l = 1$ to $9$. The values of $\tau_{0l}$ for \textit{Profile}-1 are closely aligned with those reported by \citet{BaumannAA2006}. This confirms the reliability of our analytical prescription. However, a limitation of the calculation performed by \citet{BaumannAA2006} is that the volume diffusivity ($\eta = 1\times10^{12}\,\text{cm}^2\,\text{s}^{-1}$) used to derive $\tau$ bears no direct relation to the horizontal surface diffusivity $\eta_H$ adopted in the accompanying SFT simulations. The \textit{Profile}-2 eliminates this gap, as the same physically motivated profile of $\eta(r)$ determines the decay timescale and supplies the surface value that governs horizontal transport. The dipole mode under this profile decays on a timescale of $\tau_{01} \approx 7$~yr, compared to $\approx 5$~yr for the uniform case. For both profiles, $\tau_{nl}$ decreases with increasing $l$, as higher-order modes are confined to progressively shallower depths and diffuse away more rapidly. 

\begin{table}
\centering
\caption{Decay timescales (in years) for the constant (\textit{Profile}-1) and nonuniform (\textit{Profile}-2) diffusivity profiles for spherical harmonic degrees $l=1$ to  $9$ with radial mode $n=0$.}
\label{tab:decay_times}
\begin{tabular}{ccc}
\hline\hline
$l$ & \textit{Profile}-1 & \textit{Profile}-2 \\
\hline
1 & 5.14 & 7.01  \\
2 & 4.42 & 6.05  \\
3 & 3.65 & 5.03  \\
4 & 2.96 & 4.12  \\
5 & 2.40 & 3.37  \\
6 & 1.95 & 2.79  \\
7 & 1.61 & 2.33  \\
8 & 1.34 & 2.00  \\
9 & 1.13 & 1.75  \\
\hline
\end{tabular}
\end{table}

The self-consistent estimation derived above provides the decay timescales $\tau_{nl}$ for individual modes $l$ and $n$. However, on the surface of the Sun, all spatial scales coexist, and in the polar regions the meridional flow subducts beneath the surface, contributing to the radial transport of flux, which is naturally included in the 3D STABLE model but remains absent in the 2D SFT model. In the next section, we therefore perform data-assimilated 2D SFT simulations constrained by HMI magnetogram data. The unsigned flux in the data-assimilated 2D SFT model is matched to that of the 3D data-assimilated STABLE model within $\pm 75^{\circ}$ latitude with a suitable data-assimilation interval for a consistent comparison of the resulting polar field between the two models. This comparison of polar field allows us to identify the value of $\tau\,(\equiv \tau_{0l})$ that keeps the surface dynamics of the 2D and 3D models consistent and the spatial scales to which radial decay are important.

\section{Comparison between data assimilated 2D SFT and 3D dynamo models to test estimated $\tau$}\label{sec:dataassimilatedSFT}
The eigenmode analysis in Section~\ref{sec:derivtau} provides an estimate of the radial decay timescale $\tau$ for individual spherical harmonic modes. However, on the surface of the Sun, all spatial scales coexist and interact through the nonlinear transport. To assess the effective value of $\tau$ and to determine up to which angular degree $l$ it is sufficient to retain when setting the decay term, we perform data-assimilated 2D SFT simulations in which we assimilate magnetogram data observed by HMI. For this purpose, we use the High-performance Flux Transport (HipFT) code \citep{Caplan_2025}\footnote{\url{https://github.com/predsci/hipft/tree/b05de595337ad414ff68ba3b01fed292ef8c41a1}}, an open-source 2D SFT solver that supports data assimilation in heliographic coordinates.

HipFT is the computational core of the Open-source Flux Transport (OFT) model\footnote{\url{https://github.com/predsci/OFT}}, written in Fortran, which can run in both CPU and GPU environments. It solves Equation~\ref{eq:2dsft} on a spherical grid ($\theta, \phi$) in heliographic coordinates, where $\theta$ spans from $0^{\circ}$ to $180^{\circ}$, and $\phi$ spans from $0^{\circ}$ to $360^{\circ}$. We adopt a resolution of 256 and 512 grid points along $\theta$ and $\phi$, respectively. HipFT employs a modular operator splitting implementation to compute the advection, diffusion, and source terms of Equation~\ref{eq:2dsft} at each time step. We employ an upwind and second order central difference scheme as the spatial discretizations for the advection and diffusion, respectively, with forward-Euler time integration for both. We set the horizontal diffusivity $\eta_H=3\times10^{12}\,\text{cm}^2\text{s}^{-1}$. We use the same profiles for diffusivity $\eta$ and large scale flows (meridional flow and differential rotation) in HipFT as in STABLE. We have modified the coefficients of the analytical forms used in HipFT to match the flow profiles of the 3D STABLE model. The modified profile of the meridional flow is described by 
\begin{equation}
    v_\theta(\theta) = -\left[m_1\cos\theta + m_5\cos^5\theta\right]\sin\theta,
\end{equation}
where we set $m_1=37.2~\text{ms}^{-1}$ and $m_5=-6~\text{ms}^{-1}$ to obtain peak meridional flow speed of $19~\text{ms}^{-1}$. The modified differential rotation profile is given by 
\begin{equation}
    v_\phi(\theta) = \left[d_0 + d_2\cos^2\theta + d_4\cos^4\theta\right]\sin\theta,
\end{equation}
where we set $d_0=130~\text{ms}^{-1}$, $d_2=-170~\text{ms}^{-1}$, and $d_4=-400~\text{ms}^{-1}$. 

With these flow profiles and $\eta_H$, the remaining two terms of Equation~\ref{eq:2dsft} are the source term $S_{2D}$ and the radial decay term $D(\eta)=-B_r/\tau$. We have incorporated $D(\eta)$ into the HipFT advection module for our simulations. While $D(\eta)$ accounts for the radial coupling with the interior, $S_{2D}$ provides a channel through which the observed surface radial magnetic field $B_{r, obs}$ enters in the 2D SFT model. We derive $B_{r, obs}$ from Line-Of-Sight (LOS) magnetogram data from HMI. For this purpose, we use the Magnetic Mapping and Processing (MagMAP) package from the OFT model suite to transform the LOS magnetogram data into heliographic coordinates with appropriate center-to-limb correction, and then downsample to our simulation grid to obtain $B_{r, obs}$. The weighted difference between $B_{r, obs}$ and $B_r$ is then applied as the source term
\begin{equation}
    S_{2D} = W(B_{r, obs} - B_r)/\tau_{relax},\label{eq:2dsource} 
\end{equation}
where $\tau_{relax}$ is the assimilation timescale which controls the rate at which our simulated field relaxes towards the observation, and $W$ is the weight function, which is defined as 
\begin{equation}
    W = \begin{cases}
        \mu^{\alpha} & \text{if} \ \mu > \mu_{lim} \\
        0, & \text{elsewhere}
    \end{cases}
\end{equation}
where $\mu =\cos\Theta_l$, $\Theta_l$ is the center-to-limb angle, and $\mu_{lim}$ and $\alpha$ are constants. We set $\mu_{lim}=0.1$ and $\alpha=1$. 

The weight function $W$ and the MagMAP data pipeline are shared with the 3D STABLE model \citep{Chatterjee2026ApJ}; however, the manner in which the observed data enters the model differs. In the data-assimilated 3D STABLE model, source term $\mathbf{S}_{3D}$ acts only  on the poloidal potential $C$. The MagMAP processed $B_{r,obs}$, weighted by the same function $W$ is extrapolated below the photosphere up to $0.98R_\odot$ using a hyperbolic tangent profile to obtain $B_{r,extrapol}$, and the source term $S_{3D,C}$ is then computed via spherical harmonic decomposition as 
\begin{equation}
    \tilde{S}_{3D,C} = \frac{r^2}{l(l+1)} \tilde{B}_{r,extrapol}, \label{eq:poloidal_pot_compute}
\end{equation}
where $\tilde{S}_{3D,C}$ and $\tilde{B}_{r,extrapol}$ are obtained via spherical harmonic transforms in latitude and longitude from $S_{3D,C}$ and $B_{r,extrapol}$, respectively. This procedure is inherently three-dimensional and is applied with a daily cadence during the simulation. In the 2D SFT model, no such radial structure exists, and the source term must operate entirely on the surface. We therefore use the formulation of Equation~\ref{eq:2dsource}, in which the weighted difference between the observed and simulated fields is applied as a continuous relaxation source with timescale $\tau_{relax}$. Furthermore, we restrict the assimilation within $\pm75^{\circ}$ latitude, approximately the boundary beyond which the meridional return flow begins to subduct the surface field in the 3D model. Equatorward of this boundary, the dominant mechanism coupling the surface field to the interior is radial diffusion rather than advective subduction. We calibrate $\tau_{relax}$ by matching the unsigned flux within these latitudes in our 2D SFT simulation with that of the 3D data-assimilated STABLE simulation of \cite{Chatterjee2026ApJ}, and obtain $\tau_{relax}=6$ hr. This optimum value of $\tau_{relax}$ ensures that excess flux does not accumulate due to the assimilation process. With this choice, the surface field $B_r$ evolves following Equation~\ref{eq:2dsft} under the combined action of meridional flow, horizontal diffusion and the radial decay term. Without the radial decay term, excess flux will accumulate poleward of $\pm 75^{\circ}$, as there is no meridional flow subduction to remove the surface flux in the 2D model. The proper choice of $\tau$, the only proxy of radial transport in the 2D framework, controls this accumulation of flux near the pole. 

\begin{figure*}[!htbp]
\centering
\includegraphics[width=\textwidth]{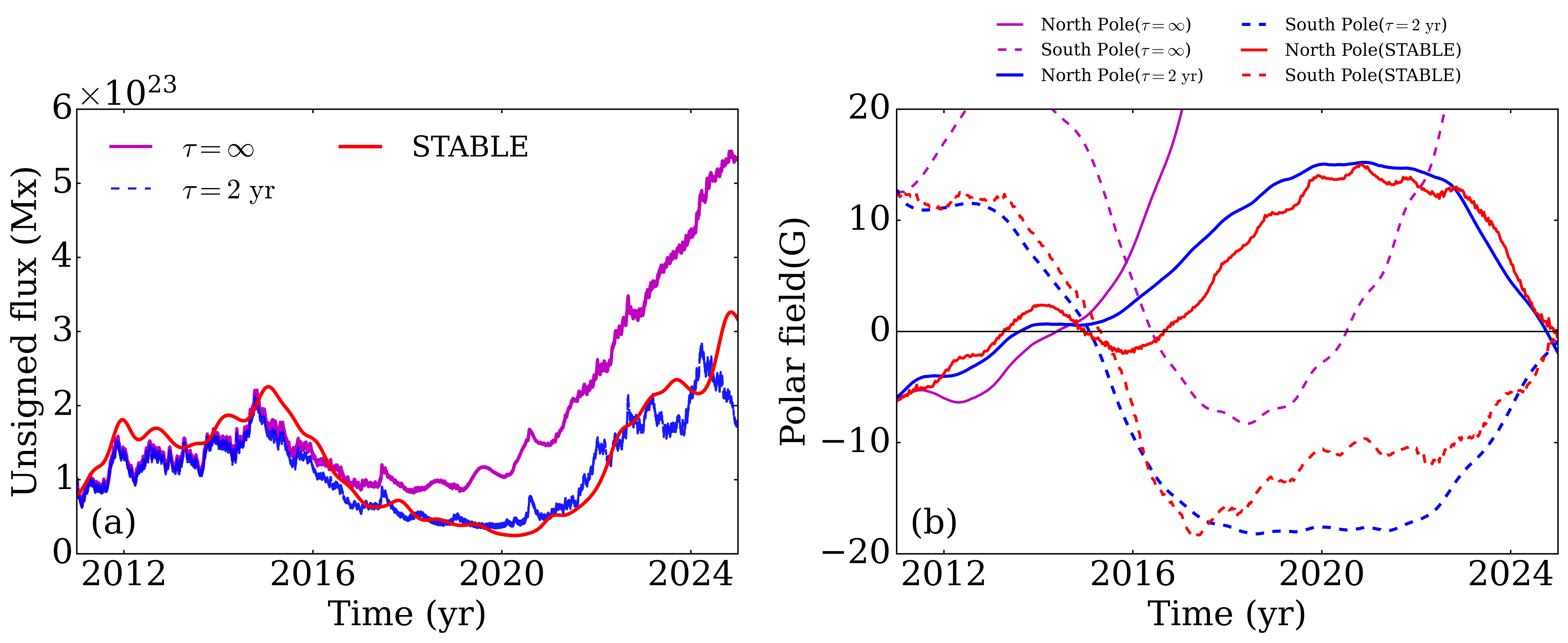}
\caption{Comparison of (a) total unsigned flux withing $\pm75^{\circ}$ latitude and (b) the polar fields (poleward $\pm75^{\circ}$) between the 2D SFT simulations and data-assimilated 3D STABLE simulation from \citet{Chatterjee2026ApJ} for the period 2011-2025. Magenta, blue, and red lines show the variation of unsigned flux in panel (a) for the 2D SFT simulation with $\tau=~\infty$, the 2D SFT simulation with $\tau=2~\text{yr}$, and the 3D STABLE model, respectively. With the same color convention, solid and dashed lines denote the north and south polar fields, respectively.}
\label{fig:da_polar_us_flux}
\end{figure*}

We perform many 2D SFT simulations with different $\tau$ to find an optimized $\tau$ for which the 2D SFT and 3D stable results would be comparable. We present here results of 2D SFT simulations with $\tau=\infty$ and $\tau = 2 ~\text{yr}$ for Cycle 24 and the rising phase of Cycle 25 (2011-2025), and compare the results with the 3D STABLE simulation. Figure~\ref{fig:da_polar_us_flux}(a) compares the unsigned flux within $\pm 75^{\circ}$ latitude from the 2D SFT simulations and the 3D STABLE model. The magenta line represents the 2D simulation with $\tau=\infty$ and red line represents the unsigned flux obtained from the 3D STABLE model. Without the radial decay term ($\tau=\infty$), the unsigned flux grows monotonically after 2018, and reaches $\sim 5.5\times 10^{23}$ Mx, as excess flux accumulates within $\pm 75 ^{\circ}$ latitude. Correspondingly, the polar field (poleward of $\pm 75 ^{\circ}$), shown in Figure~\ref{fig:da_polar_us_flux}(b) with solid and dashed lines for the north and south poles, respectively (magenta for $\tau=\infty$, red for STABLE), grows to a high amplitude ($> 20 $G) over the same period. Since we use the same flow profiles ($v_\theta$, $v_{\phi}$) and surface diffusivity $\eta_H$ as in the 3D STABLE model, this excess is entirely attributable to the absence of the $D(\eta)$ and meridional flow subduction in the 2D SFT model;the poleward transported flux accumulates indefinitely at the poles with no mechanism for removal. 

\begin{figure*}[t]
\centering
\includegraphics[width=0.9\textwidth]{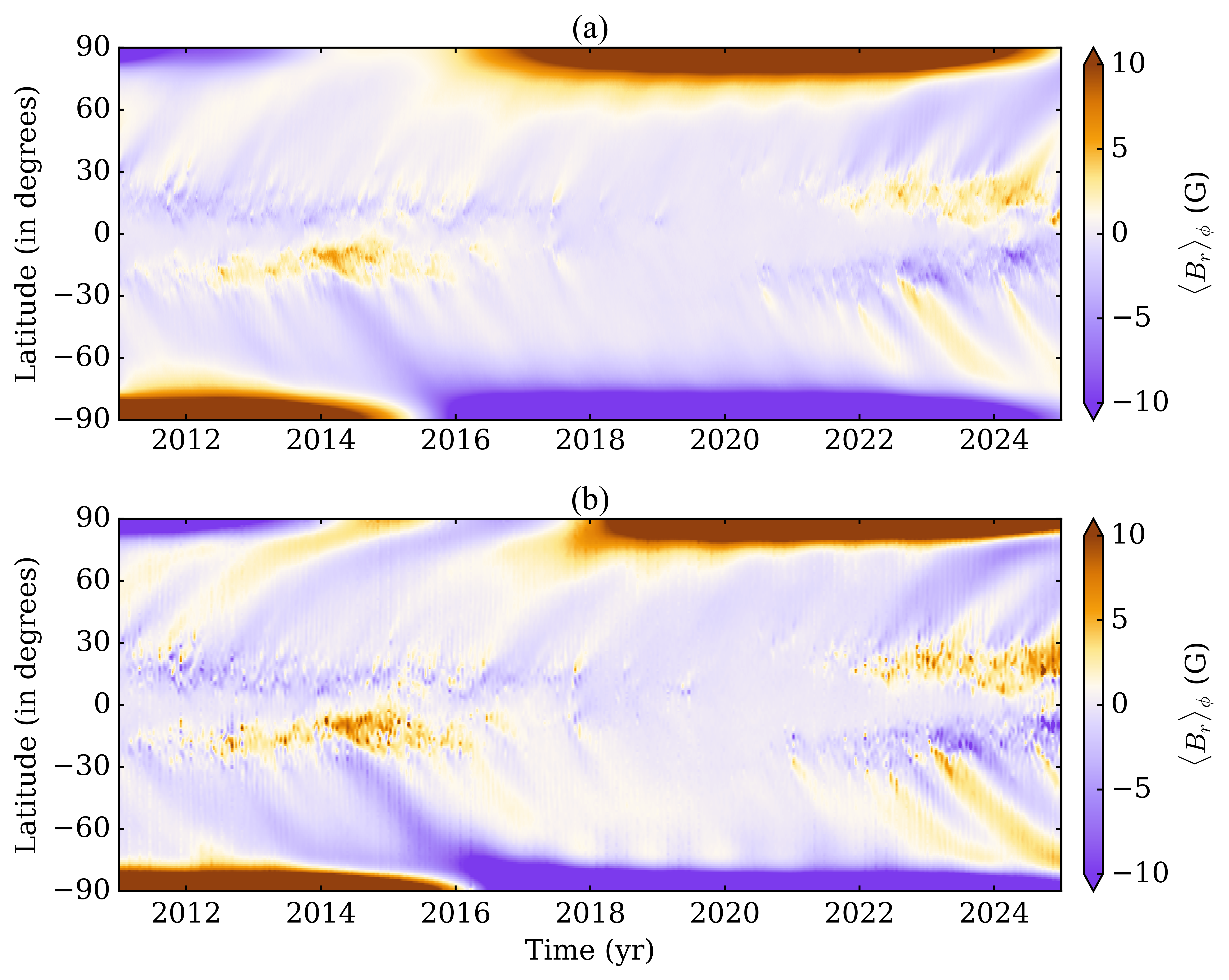}
\caption{Longitudinally averaged radial magnetic field $\langle B_r\rangle_\phi$ as a function of latitude and time - the butterfly diagram - for (a) the 2D SFT simulation with daily magnetogram assimilation and $\tau=2~\text{yr}$, and (b) the 3D STABLE model with daily magnetogram assimilation from \citet{Chatterjee2026ApJ}, spanning Cycle 24 and the rising phase of Cycle 25 (2011 - 2025 yr).}
\label{fig:da_bfly}
\end{figure*}

\begin{figure*}[!htbp]
\centering
\includegraphics[width=0.9\textwidth]{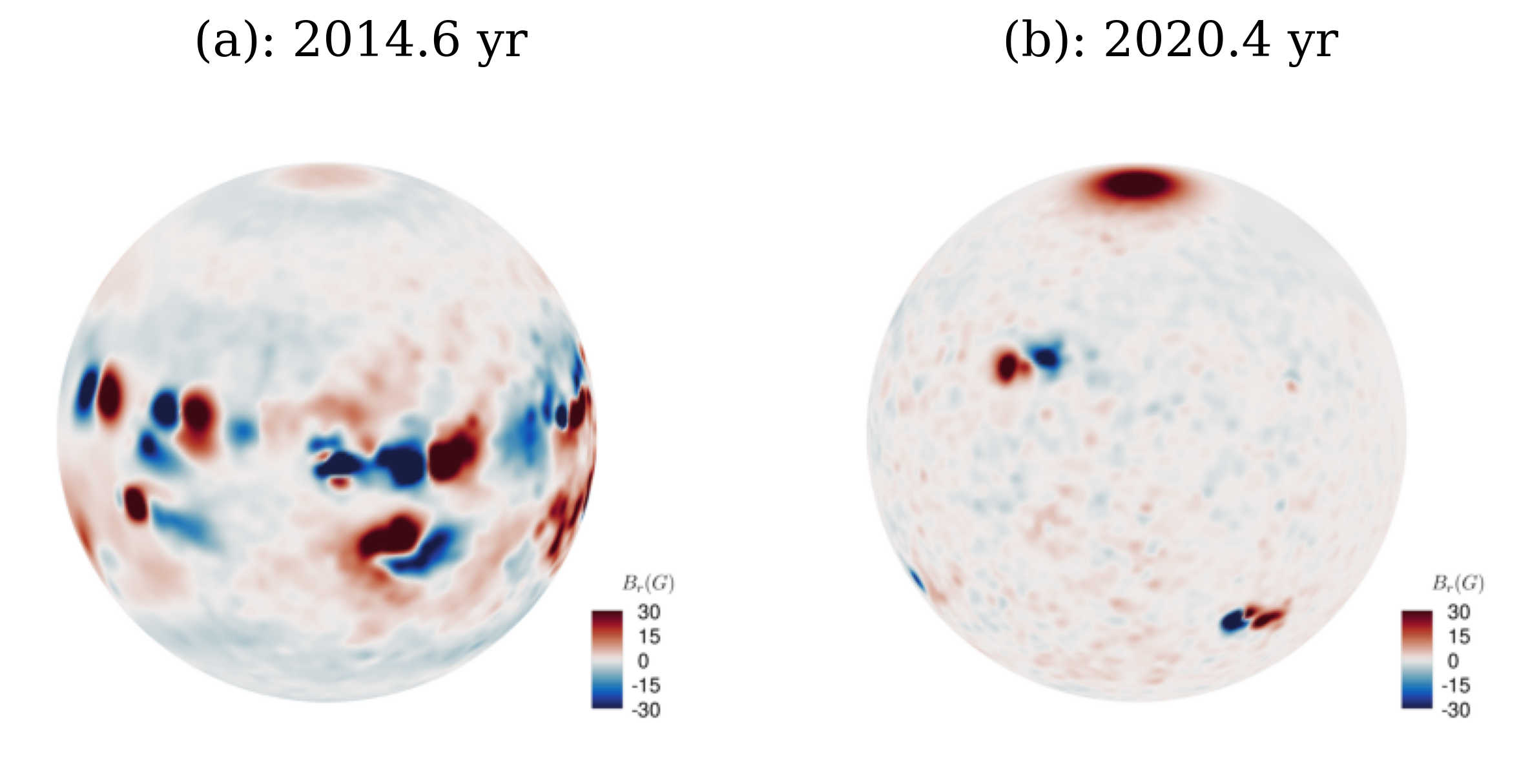}
\caption{Snapshots of the radial magnetic field $B_r$ on the solar surface obtained from the 3D data-assimilated STABLE simulation \citep{Chatterjee2026ApJ} at (a) 2014.6 yr, near the maximum of Cycle 24, and (b) 2020.4 yr, during the minimum between Cycles 24 and 25.}
\label{fig:da_3d_snap}
\end{figure*}

In contrast, the simulation with $\tau=2~\text{yr}$ (blue solid line) closely tracks the unsigned flux of the 3D STABLE simulation, as shown in Figure~\ref{fig:da_polar_us_flux}(a). In the Figure~\ref{fig:da_polar_us_flux}(b), the evolution of polar fields are shown with blue solid and dashed lines, which are in close agreement with the corresponding red lines from the 3D STABLE simulation. The polarity reversal around 2014-2015 yr and the subsequent build up of polar field for the Cycle 25 are well reproduced. Figure~\ref{fig:da_bfly} further compares the longitudinally averaged $\langle B_r\rangle_\phi$ as a function of latitude and time (the butterfly diagram) for (a) the 2D SFT simulation with $\tau=2~\text{yr}$ and (b) the 3D STABLE model. Both the butterfly diagrams, exhibit the characteristic equatorward migration of active region flux, the poleward transport of trailing-polarity flux, and the polar field reversal around 2014–2016 yr. Some finer structures, such as weak poleward surge in the northern and southern hemispheres, and distribution of alternating positive and negative flux regimes above $60^{\circ}$ after the reversal, are not fully reproduced in the 2D simulation, as a single decay timescale cannot capture the mode-dependent radial diffusion of the 3D model; nevertheless, the large-scale features are well reproduced. Referring to Table~\ref{tab:decay_times}, $\tau = 2 ~\text{yr}$ corresponds to the decay timescale of $l=8$ mode, rather than the slowest decaying dipole mode ($l=1$, $\tau \sim 7~\text{yr}$), when the radial diffusivity $\eta$ follows \textit{Profile-2} as used in the 3D STABLE model \citep{Chatterjee2026ApJ}, for comparison. This is because the single $\tau$ must account for the radial decay of all modes simultaneously present on the surface, as the magnetogram assimilation continuously replenishes all scales at each assimilation step during the simulation. Figure~\ref{fig:da_3d_snap} shows two snapshots of $B_r$ on the solar surface from the 3D STABLE simulation near the maximum of Cycle 24 and during the minimum between Cycles 24 and 25, illustrating the simultaneous presence of multiple spatial scales on the solar surface. If $\tau$ is set to the slowest dipole decay time scale, higher order modes will take longer time to decay than they would in the 3D model, leading to excess flux accumulation at the poles. 

In the next section, we set up a conventional 2D SFT simulation with discrete large-scale active regions as the source term $S_{2D}$ without any small-scale diffuse field to demonstrate whether the $\tau$ as derived in Section~\ref{sec:derivtau} corresponding to large scale can reproduce observed behavior of surface field without any secular drift problem in polar field for a given value of $\eta_H$ and radial profile inside the convection zone. In the traditional SFT setting, $\eta_H$ removes the small scales generated from the active regions, and the polar field is built by the contribution from the large-scale dipole component only. The effective $\tau$ therefore recovers the dipole value $\tau_{01} \sim 7~\text{yr}$ from Table~\ref{tab:decay_times}.

\section{Can the estimated decay time scale corresponding to large-scale ($\tau_{01}$) solve secular drift problem in traditional 2D SFT model with only active regions? }\label{sec:test_w_SFT_sharp}
   
%\section{Testing the estimated $\tau_{01}$ as a solution of secular drift problem with 2D SFT model}\label{sec:test_w_SFT_sharp}

In this section, we replace daily magnetogram assimilation with active regions as the source term $S_{2D}$ in Equation~\ref{eq:2dsft}. We use the 2D SFT model of \citet{YeatesSoPh2015} and have translated the code into Python for this work. The code solves Equation~\ref{eq:2dsft} in terms of the vector potential $[A_\theta, A_\phi]$\citep{YeatesSoPh2015, VirtanenAA2017}, from which $B_r$ is recovered via 
\begin{equation}
    B_r = \frac{1}{R_\odot\sin\theta}\left[\frac{\partial(\sin\theta\, A_\phi)}{\partial\theta} - \frac{\partial A_\theta}{\partial\phi}\right].
    \label{eq:vecpot}
\end{equation}
This formulation guarantees that magnetic flux is automatically conserved throughout the simulation, including after the injection of active regions as the source term. This is in contrast to HipFT (Section~\ref{sec:dataassimilatedSFT}), which directly solves $B_r$ and requires that flux conservation be enforced explicitly. 

We extract the active regions from synoptic magnetograms observed by the Michelson Doppler Imager (MDI; Cycle 23) and HMI (Cycles 24 and 25). This dataset spans Carrington Rotations 1950 to 2297 (1999 - 2025 yr). For each Carrington rotation, the magnetogram is smoothed with a Gaussian filter ($\sigma = 3 $ grid cells), and pixels at which the absolute value of $B_r$ exceeds a threshold of 40~G are grouped into connected regions. These identified regions whose net flux imbalance exceeds half of the total unsigned flux are discarded as unipolar remnants rather than newly emerged bipolar active regions. Each surviving region is corrected for flux balance and inserted into the simulation (as $S_{2D}$) at the time its centroid crosses the central meridian. A local flux correction is applied during insertion to preserve the total signed flux in the simulation domain. Note that in this way, each region preserves the morphological complexity of the original active region, a feature that would be lost if the regions were approximated as idealized bipolar magnetic regions (BMRs). For further details, see \citet{YeatesSoPh2015} and \citet{VirtanenAA2017}.

\begin{figure*}[!htbp]
\centering
\includegraphics[width=\textwidth]{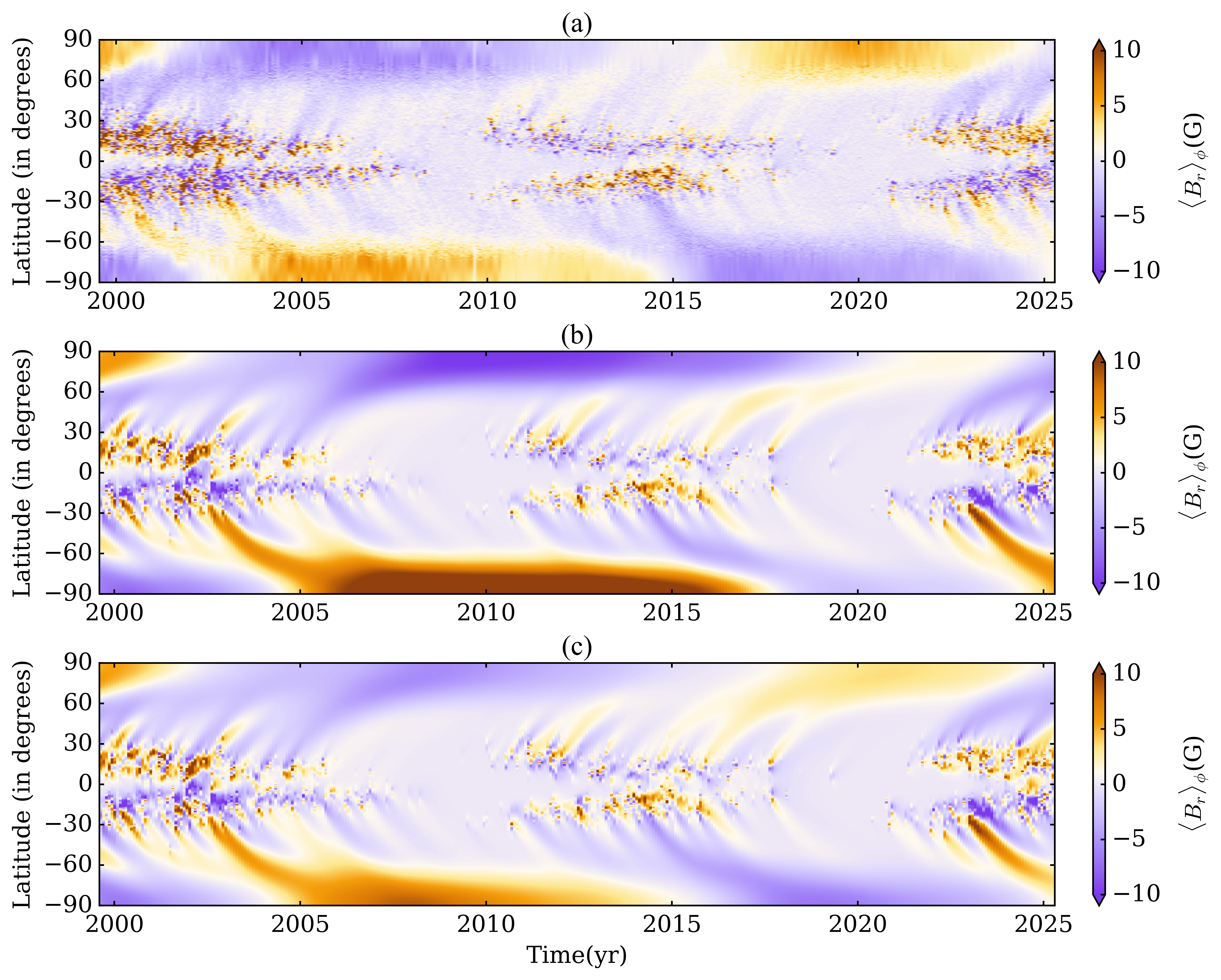}
\caption{Magnetic butterfly diagrams (longitude-averaged $\langle B_r\rangle _\phi$) for approximately 26 years (1999-2025.3 yr). Panel (a): observed butterfly diagram constructed from MDI (CR 1950–2104) and HMI (CR 2096–2297) synoptic magnetograms. Panel (b): 2D SFT simulation driven by discrete active regions with no radial decay term ($\tau=\infty$). Panel (c): same as (b) but with the dipole decay timescale ($\tau=7~\text{yr}$). 
}
\label{fig:butterfly_sft_yeates}
\end{figure*}

Unlike Section~\ref{sec:dataassimilatedSFT}, where the same flow profiles as the 3D STABLE model were used in 2D SFT to isolate the effect of $\tau$, here we adopt the standard transport parameters commonly used in the SFT community. Our aim here is not to compare 2D SFT results with the 3D model anymore; rather, we want to verify whether the analytically estimated $\tau_{01}$ can remove the secular drift problem and whether it is an accurate decay timescale for large-scale surface fields. For differential rotation, we use the Snodgrass–Ulrich profile in the Carrington frame \citep{Snodgrass1990ApJ}:
    \label{eq:dr_yeates}
\begin{equation}
    \omega(\theta) = \left(0.521 - 2.396\cos^2\theta - 1.787\cos^4\theta\right)~\text{deg day}^{-1},
\end{equation}
where $\omega$ is the angular velocity and $v_\phi = R_\odot\omega(\theta)\sin\theta$. For the meridional flow, we use the semi-empirical profile of \citet{Schssler2006AAP}:
\begin{equation}
v_\theta(\theta) = v_0 \sin\left(2\left(\frac{\pi}{2} - \theta\right)\right) \exp\left(\pi - 2\left|\frac{\pi}{2} - \theta\right|\right),
\label{eq:mf_yeates}
\end{equation}
with a peak velocity of $v_0=11 ~ ms^{-1}$ \citep{YeatesSoPh2015, VirtanenAA2017}. %Note that here $v_0$ differs from the peak amplitude of the meridional flow ($\sim 19~\text{ms}^{-1}$) used in the  Section.~\ref{sec:dataassimilatedSFT}, where the flow was matched to the 3D STABLE model.
The slower flow ensures that the spread of accumulated flux is visible from $55^{\circ}-60^{\circ}$ latitude, which is consistent with the butterfly diagram obtained from the observed synoptic maps in Figure.~\ref{fig:butterfly_sft_yeates}(a). This amplitude lies within the observed range of ~$10-20~\text{ms}^{-1}$ \citep{HanasogeLRSP2022}. The horizontal diffusivity $\eta_H$ is set to $\eta_H=3\times10^{12}\,\text{cm}^2\text{s}^{-1}$ as in Section~\ref{sec:dataassimilatedSFT}. The computational grid is uniform in sin(latitude) and longitude with a resolution of $90\times180$. We use a forward Euler time integration scheme, with a timestep of $\Delta t=0.2$days determined by the Courant–Friedrichs–Lewy (CFL) condition.  

With the above setup, we perform two simulations: one with $\tau=\infty$ and one with $\tau=7 ~\text{yr}$. Since active regions are the sole source of magnetic flux, horizontal diffusion erodes the small-scale structure, leaving only the large-scale dipole component to reach the polar caps, the effective decay timescale in this regime is therefore expected to recover the dipole value $\tau = \tau_{01}\sim7~\text{yr}$ using eigenmode analysis from Table~\ref{tab:decay_times}. Figure~\ref{fig:butterfly_sft_yeates} shows the butterfly diagrams obtained for the Cycles 23-25 from the observed synoptic magnetograms of MDI and HMI (panel a), the SFT simulation with $\tau=\infty$ (panel b), and the SFT simulation with $\tau=7~\text{yr}$ (panel c). The equatorward migration of active region flux is reproduced in both simulations. The key difference lies in the polar field evolution (poleward of $\pm65^{\circ}$). Without the decay term (panel b), the accumulated polar fields\ in the southern hemisphere  after the Cycle 23 stays persists for an extended period and reverses only after $\sim$2016 yr due to a strong negative poleward surge. In the northern hemisphere, the polar field generated during Cycle 23 nearly remains throughout the weak Cycle 24 and no sufficiently strong poleward surge is present to reverse it. This is the well-known secular drift problem in the classical SFT model without radial decay term \citep{Schrijver2002ApJ, BaumannAA2006, VirtanenAA2017}. With $\tau=7~\text{yr}$ (panel c), the butterfly diagram reproduces the observed large-scale magnetic features. The poleward surges of the trailing polarity flux are clearly visible, and the polar fields reverse at the same times as observed. 

\begin{figure}[!htbp]
\centering
\includegraphics[width=0.48\textwidth]{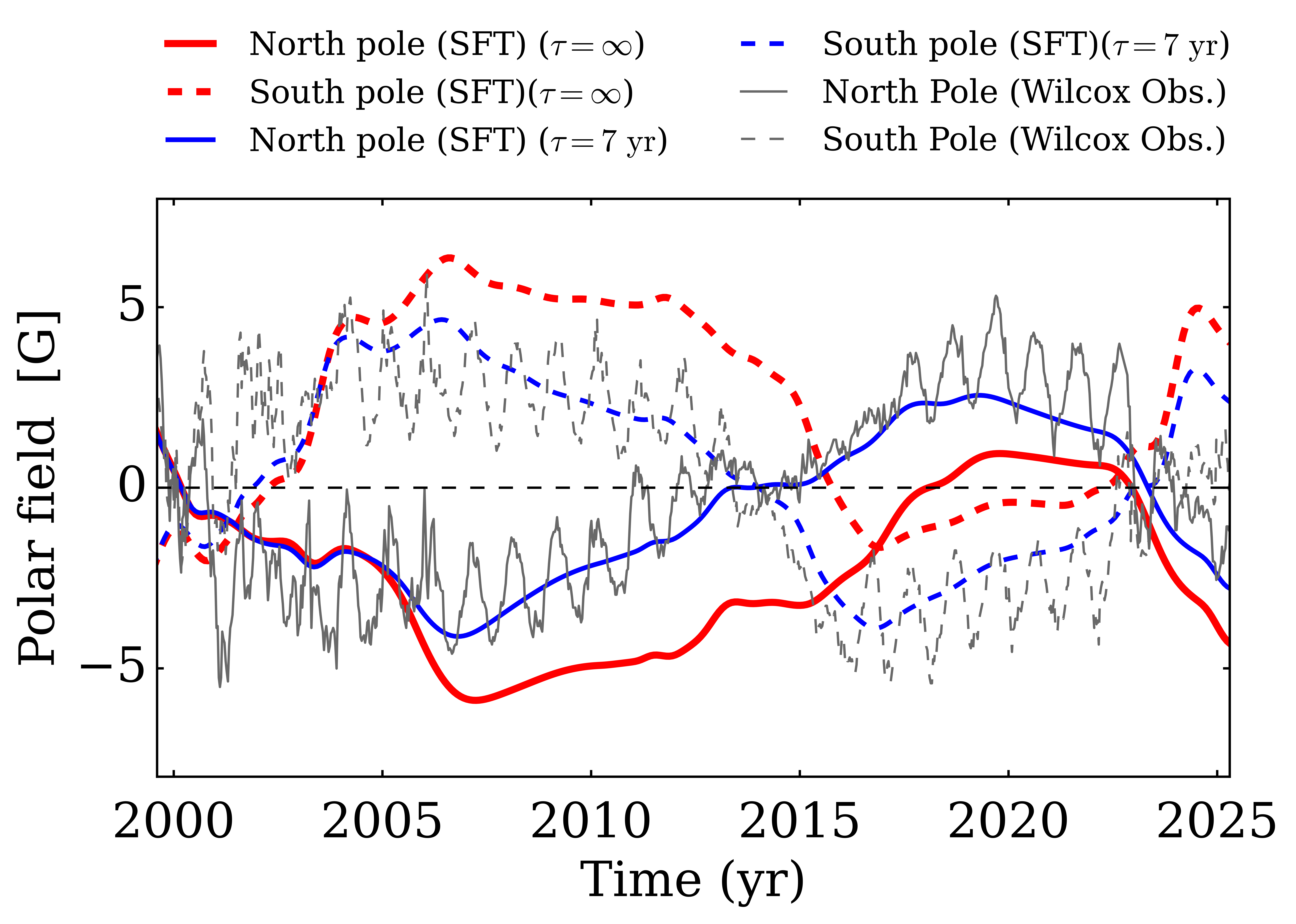}
\caption{Evolution of the polar fields (poleward of $\pm55^{\circ}$). Gray curves: WSO observations (solid: north, dashed: south). Red curves: SFT simulation with $\tau = \infty$. Blue curves: SFT simulation with $\tau = 7~yr$.}
\label{fig:polar_fields_sft_yeates}
\end{figure}

We compare the polar field (poleward of $\pm55^{\circ}$) evolution from both the simulations with the polar field measurements from the Wilcox Solar Observatory (WSO). In Figure~\ref{fig:polar_fields_sft_yeates}, gray solid and dashed curves represent the observed polar field variation from WSO. Without the decay term ($\tau=\infty$, red solid and dashed curves), the polar field amplitude exceeds 5~G and the subsequent Cycle 24 reversal is delayed in both hemispheres. With $\tau=7~\text{yr}$ (blue solid and dashed curves), the polar fields closely follow the WSO observations in terms of amplitudes and reversal times. In addition, the subsequent buildup of the Cycle 25 is also well captured.  The agreement confirms that the radial decay timescale recovers the dipole value $\tau \sim 7~\text{yr}$ when only discrete active regions serve as the source term. 
 
\section{Conclusions}\label{sec:conclusion}
In summary, we have presented a systematic, physics-driven analysis of the radial decay timescale $\tau$ in the 2D SFT model. We derived $\tau$ self-consistently from the induction equation in a 3D mean-field framework STABLE using a depth-dependent turbulent diffusivity profile ($\eta(r)$) for the convection zone. This includes the physical origin of $\tau$ due to the radial transport of magnetic flux in the convection zone. Keeping the same surface transport parameters as in the 3D model, we derived an eigenvalue equation by treating the coupling between the surface and the interior as a diffusion problem in a 3D spherical shell. The solutions of the eigenvalue equation provide the decay timescale $\tau_{nl}$ for each angular and radial mode $l$ and $n$ for a given $\eta(r)$. The resulting spectrum of scale-dependent radial decay timescales decrease with increasing spherical harmonic degree $l$ for $n=0$. For uniform $\eta(r)$, our results match the earlier estimates of $\eta(r)$ by \cite{BaumannAA2006}, while for a physically motivated two-step profile of $\eta(r)$, the slowest decaying mode is the dipole mode ($l=1$), with a decay timescale of $\sim7$ yr. This derivation serves as an alternative to the conventional approach of treating $\tau$ as a free parameter and optimizing against surface observations.  

We validated the derived spectrum, $\tau_{nl}$ through two complementary numerical experiments, since the subduction of flux by the meridional flow near the pole (poleward $\sim 75^{\circ})$ also plays a significant role alongside radial diffusion in transporting flux in the convection zone. First, we assimilated daily HMI magnetograms into the 2D SFT model within $\pm75^{\circ}$ latitude from 2011 to 2025 and compared the polar fields poleward of $\pm75^{\circ}$ latitude with those of the 3D data-assimilated STABLE model, using the same surface diffusivity, meridional flow, and differential rotation. We found that modes up to spherical harmonic degree $l=8$ contribute to polar field formation comparable to that of the 3D model, while modes beyond this degree decay before reaching the polar caps. The corresponding $\tau_{nl}$ for $l=8$ is $\sim 2$ yr according to Table~\ref{tab:decay_times}. In contrast, when only discrete active regions drive the traditional 2D SFT model for Cycles 23-25, without any small scale field (remnant from active regions also smoothen out by horizontal diffusivity) and polar fields is built by the large-scale dipole component alone; we find the effective $\tau=\tau_{01}\sim 7$yr is accurately reproducing the polar field and surface field dynamics according to observation. SFT model with this value removes the secular drift problem associated with the weak Cycle 24 following the strong Cycle 23, and the reproduced polar field is consistent with WSO observations.  

This is, to our knowledge, the first self-consistent derivation of $\tau$ from a physically motivated prescription tied to radial transport of flux from the surface in the 3D model. Our findings have direct implications for century-long reconstruction of the solar polar magnetic field and for predicting the next cycle amplitude based on the polar field of the current cycle, as these analyses depend on the accurate generation of polar field with the 2D SFT model. Our eigenmode analysis framework is general and can provide $\tau$ for a different value of surface diffusivity with the same profile or a different radial profile of turbulent diffusivity, making it applicable beyond the specific profile adopted here. A natural extension of this work is to explore how $\tau$ responds to different diffusivity profiles, surface diffusivity values, and meridional flow profile to assess the sensitivity of the estimated $\tau$ to the cycle dependent transport parameters; we leave this for future work.
  
%% Please use the acknowledgment and contribution environments. This will 
%% be anonomyized when the "anonymous" style option is used. 
\begin{acknowledgments}
%The authors would like to thank two anonymous referees for their constructive comments that helped a lot to improve our manuscript. 
The computations were performed on our local server - Parker, and the Pegasus cluster at IUCAA Pune. Access to the Pegasus cluster was obtained through Project hpc2601014. Some runs were performed on the Param Sanganak supercomputer at IIT Kanpur. Access to Param Sanganak was provided by the National Supercomputing Mission and Indian Institute of Technology Kanpur. The authors thank Prof. H M Antia for kindly providing the differential rotation data for the last three decades. The HMI data used are courtesy of the NASA/SDO and the HMI science team. We used MDI data from SOHO, a project of international cooperation between ESA and NASA. We thank Lisa Upton, and Bibhuti Jha for useful discussions. The authors acknowledge funding from the IIT Kanpur initiation grant IITK/PHY/2022386 and special grant IITK/PHY/2025623. 

\end{acknowledgments}

\facilities{SOHO(MDI), SDO(HMI)}%, AAVSO, CTIO:1.3m, CTIO:1.5m, CXO}

\software{sft\_data\citep{sftyeates}, HipFT, MagMAP \citep{MAGMAP}}%astropy \citep{2013A&A...558A..33A,2018AJ....156..123A,2022ApJ...935..167A},  
          %Cloudy \citep{2013RMxAA..49..137F}, 
          %Source Extractor \citep{1996A&AS..117..393B}
          %}


%\appendix

%\section{}\label{sec:source term}

%% For this sample we use BibTeX plus aasjournalv7.bst to generate the

%%\bibliography{sample701}{}
\bibliographystyle{aasjournalv7}
\bibliography{ms, our_ref}

@misc{MAGMAP,
  author       = {{Predictive Science Inc.}},
  year         = {2024},
  title        = {MAGMAP: Magnetic Mapping Pipeline},
  howpublished = {\url{https://github.com/predsci/magmap}},
  note         = {[accessed: May 2025]}
}

@article{Caplan_2025,
doi = {10.3847/1538-4365/adc080},
year = {2025},
month = {may},
publisher = {The American Astronomical Society},
volume = {278},
number = {1},
pages = {24},
author = {Caplan, Ronald M. and Stulajter, Miko M. and Linker, Jon A. and Downs, Cooper and Upton, Lisa A. and Jha, Bibhuti Kumar and Attie, Raphael and Arge, Charles N. and Henney, Carl J.},
title = {Open-source Flux Transport (OFT). I. HipFT–High-performance Flux Transport},
journal = {The Astrophysical Journal Supplement Series},
}

@ARTICLE{Miesch_Dikpati_2014,
       author = {{Miesch}, Mark S. and {Dikpati}, Mausumi},
        title = "{A Three-dimensional Babcock-Leighton Solar Dynamo Model}",
      journal = {The Astrophysical Journal Letters},
         year = 2014,
        month = apr,
       volume = {785},
       number = {1},
          eid = {L8},
        pages = {L8},
          doi = {10.1088/2041-8205/785/1/L8},
archivePrefix = {arXiv},
       eprint = {1401.6557},
 primaryClass = {astro-ph.SR},
       adsurl = {https://ui.adsabs.harvard.edu/abs/2014ApJ...785L...8M}
}

@article{Miesch_Teweldebirhan_2016,
title = {A three-dimensional Babcock–Leighton solar dynamo model: Initial results with axisymmetric flows},
journal = {Advances in Space Research},
volume = {58},
number = {8},
pages = {1571-1588},
year = {2016},
note = {Solar Dynamo Frontiers},
issn = {0273-1177},
doi = {https://doi.org/10.1016/j.asr.2016.02.018},
author = {Mark S. Miesch and Kinfe Teweldebirhan}
}

@article{Hazra_2017,
doi = {10.3847/1538-4357/835/1/39},
url = {https://dx.doi.org/10.3847/1538-4357/835/1/39},
year = {2017},
month = {jan},
publisher = {The American Astronomical Society},
volume = {835},
number = {1},
pages = {39},
author = {Hazra, Gopal and Choudhuri, Arnab Rai and Miesch, Mark S.},
title = {A THEORETICAL STUDY OF THE BUILD-UP OF THE SUN’S POLAR MAGNETIC FIELD BY USING A 3D KINEMATIC DYNAMO MODEL},
journal = {The Astrophysical Journal}
}

@ARTICLE{Upton_2014ApJ,
       author = {{Upton}, Lisa and {Hathaway}, David H.},
        title = "{Predicting the Sun's Polar Magnetic Fields with a Surface Flux Transport Model}",
      journal = {\apj},
         year = 2014,
        month = jan,
       volume = {780},
       number = {1},
          eid = {5},
        pages = {5},
          doi = {10.1088/0004-637X/780/1/5},
archivePrefix = {arXiv},
       eprint = {1311.0844},
 primaryClass = {astro-ph.SR},
       adsurl = {https://ui.adsabs.harvard.edu/abs/2014ApJ...780....5U}
}

@ARTICLE{Jiang2014SSRv,
       author = {{Jiang}, J. and {Hathaway}, D.~H. and {Cameron}, R.~H. and {Solanki}, S.~K. and {Gizon}, L. and {Upton}, L.},
        title = "{Magnetic Flux Transport at the Solar Surface}",
      journal = {Space Science Reviews},
         year = 2014,
        month = dec,
       volume = {186},
       number = {1-4},
        pages = {491-523},
          doi = {10.1007/s11214-014-0083-1},
archivePrefix = {arXiv},
       eprint = {1408.3186},
 primaryClass = {astro-ph.SR},
       adsurl = {https://ui.adsabs.harvard.edu/abs/2014SSRv..186..491J}
}

@ARTICLE{Lemerle_2015ApJ,
       author = {{Lemerle}, Alexandre and {Charbonneau}, Paul and {Carignan-Dugas}, Arnaud},
        title = "{A Coupled 2 {\texttimes} 2D Babcock-Leighton Solar Dynamo Model. I. Surface Magnetic Flux Evolution}",
      journal = {\apj},
         year = 2015,
        month = sep,
       volume = {810},
       number = {1},
          eid = {78},
        pages = {78},
          doi = {10.1088/0004-637X/810/1/78},
archivePrefix = {arXiv},
       eprint = {1511.08548},
 primaryClass = {astro-ph.SR},
       adsurl = {https://ui.adsabs.harvard.edu/abs/2015ApJ...810...78L}
}

@ARTICLE{Yeates_2023SSRv,
       author = {{Yeates}, Anthony R. and {Cheung}, Mark C.~M. and {Jiang}, Jie and {Petrovay}, Kristof and {Wang}, Yi-Ming},
        title = "{Surface Flux Transport on the Sun}",
      journal = {\ssr},
         year = 2023,
        month = jun,
       volume = {219},
       number = {4},
          eid = {31},
        pages = {31},
          doi = {10.1007/s11214-023-00978-8},
archivePrefix = {arXiv},
       eprint = {2303.01209},
 primaryClass = {astro-ph.SR},
       adsurl = {https://ui.adsabs.harvard.edu/abs/2023SSRv..219...31Y}
}

@ARTICLE{Cameron_2010ApJ,
       author = {{Cameron}, R.~H. and {Jiang}, J. and {Schmitt}, D. and {Sch{\"u}ssler}, M.},
        title = "{Surface Flux Transport Modeling for Solar Cycles 15-21: Effects of Cycle-Dependent Tilt Angles of Sunspot Groups}",
      journal = {\apj},
         year = 2010,
        month = aug,
       volume = {719},
       number = {1},
        pages = {264-270},
          doi = {10.1088/0004-637X/719/1/264},
archivePrefix = {arXiv},
       eprint = {1006.3061},
 primaryClass = {astro-ph.SR},
       adsurl = {https://ui.adsabs.harvard.edu/abs/2010ApJ...719..264C}
}

@ARTICLE{Dash_2024ApJ,
       author = {{Dash}, Soumyaranjan and {DeRosa}, Marc L. and {Dikpati}, Mausumi and {Sun}, Xudong and {Mahajan}, Sushant S. and {Liu}, Yang and {Hoeksema}, J. Todd},
        title = "{Ensemble Kalman Filter Data Assimilation into the Surface Flux Transport Model to Infer Surface Flows: An Observing System Simulation Experiment}",
      journal = {\apj},
         year = 2024,
        month = nov,
       volume = {975},
       number = {2},
          eid = {288},
        pages = {288},
          doi = {10.3847/1538-4357/ad7eac},
archivePrefix = {arXiv},
       eprint = {2409.15233},
 primaryClass = {astro-ph.SR},
       adsurl = {https://ui.adsabs.harvard.edu/abs/2024ApJ...975..288D}
}

@ARTICLE{Schrijver_2003SoPh,
       author = {{Schrijver}, Carolus J. and {De Rosa}, Marc L.},
        title = "{Photospheric and heliospheric magnetic fields}",
      journal = {\solphys},
         year = 2003,
        month = jan,
       volume = {212},
       number = {1},
        pages = {165-200},
          doi = {10.1023/A:1022908504100},
       adsurl = {https://ui.adsabs.harvard.edu/abs/2003SoPh..212..165S}
}

@ARTICLE{Hickmann_2015SoPh,
       author = {{Hickmann}, Kyle S. and {Godinez}, Humberto C. and {Henney}, Carl J. and {Arge}, C. Nick},
        title = "{Data Assimilation in the ADAPT Photospheric Flux Transport Model}",
      journal = {\solphys},
         year = 2015,
        month = apr,
       volume = {290},
       number = {4},
        pages = {1105-1118},
          doi = {10.1007/s11207-015-0666-3},
archivePrefix = {arXiv},
       eprint = {1410.6185},
 primaryClass = {math-ph},
       adsurl = {https://ui.adsabs.harvard.edu/abs/2015SoPh..290.1105H}
}

@ARTICLE{Jiang_2023,
       author = {{Jiang}, Jie and {Zhang}, Zebin and {Petrovay}, Krist{\'o}f},
        title = "{Comparison of physics-based prediction models of solar cycle 25}",
      journal = {Journal of Atmospheric and Solar-Terrestrial Physics},
         year = 2023,
        month = feb,
       volume = {243},
          eid = {106018},
        pages = {106018},
          doi = {10.1016/j.jastp.2023.106018},
archivePrefix = {arXiv},
       eprint = {2212.01158},
 primaryClass = {astro-ph.SR},
       adsurl = {https://ui.adsabs.harvard.edu/abs/2023JASTP.24306018J}
}

@ARTICLE{MackayLRSP2012,
       author = {{Mackay}, Duncan H. and {Yeates}, Anthony R.},
        title = "{The Sun's Global Photospheric and Coronal Magnetic Fields: Observations and Models}",
      journal = {Living Reviews in Solar Physics},
         year = 2012,
        month = dec,
       volume = {9},
       number = {1},
          eid = {6},
        pages = {6},
          doi = {10.12942/lrsp-2012-6},
archivePrefix = {arXiv},
       eprint = {1211.6545},
 primaryClass = {astro-ph.SR},
       adsurl = {https://ui.adsabs.harvard.edu/abs/2012LRSP....9....6M}
}

@ARTICLE{SchrijverSoPh2003,
       author = {{Schrijver}, Carolus J. and {De Rosa}, Marc L.},
        title = "{Photospheric and heliospheric magnetic fields}",
      journal = {\solphys},
         year = 2003,
        month = jan,
       volume = {212},
       number = {1},
        pages = {165-200},
          doi = {10.1023/A:1022908504100},
       adsurl = {https://ui.adsabs.harvard.edu/abs/2003SoPh..212..165S}
}

@ARTICLE{JiangJASTP2023,
       author = {{Jiang}, Jie and {Zhang}, Zebin and {Petrovay}, Krist{\'o}f},
        title = "{Comparison of physics-based prediction models of solar cycle 25}",
      journal = {Journal of Atmospheric and Solar-Terrestrial Physics},
         year = 2023,
        month = feb,
       volume = {243},
          eid = {106018},
        pages = {106018},
          doi = {10.1016/j.jastp.2023.106018},
archivePrefix = {arXiv},
       eprint = {2212.01158},
 primaryClass = {astro-ph.SR},
       adsurl = {https://ui.adsabs.harvard.edu/abs/2023JASTP.24306018J}
}

@ARTICLE{LeightonApJ1964,
       author = {{Leighton}, Robert B.},
        title = "{Transport of Magnetic Fields on the Sun.}",
      journal = {\apj},
         year = 1964,
        month = nov,
       volume = {140},
        pages = {1547},
          doi = {10.1086/148058},
       adsurl = {https://ui.adsabs.harvard.edu/abs/1964ApJ...140.1547L}
}

@ARTICLE{WangSci1989,
       author = {{Wang}, Y. M. and {Nash}, A.~G. and {Sheeley}, Jr., N.~R.},
        title = "{Magnetic Flux Transport on the Sun}",
      journal = {Science},
         year = 1989,
        month = aug,
       volume = {245},
       number = {4919},
        pages = {712-718},
          doi = {10.1126/science.245.4919.712},
       adsurl = {https://ui.adsabs.harvard.edu/abs/1989Sci...245..712W}
}

@ARTICLE{SheeleyLRSP2005,
       author = {{Sheeley}, Jr., Neil R.},
        title = "{Surface Evolution of the Sun's Magnetic Field: A Historical Review of the Flux-Transport Mechanism}",
      journal = {Living Reviews in Solar Physics},
         year = 2005,
        month = dec,
       volume = {2},
       number = {1},
          eid = {5},
        pages = {5},
          doi = {10.12942/lrsp-2005-5},
       adsurl = {https://ui.adsabs.harvard.edu/abs/2005LRSP....2....5S}
}

@ARTICLE{DeVoreSoPh1984,
       author = {{DeVore}, C.~R. and {Boris}, J.~P. and {Sheeley}, Jr., N.~R.},
        title = "{The concentration of the large-scale solar magnetic field by a meridional surface flow}",
      journal = {\solphys},
         year = 1984,
        month = may,
       volume = {92},
       number = {1-2},
        pages = {1-14},
          doi = {10.1007/BF00157230},
       adsurl = {https://ui.adsabs.harvard.edu/abs/1984SoPh...92....1D}
}

@ARTICLE{WangApJ1991,
       author = {{Wang}, Y.-M. and {Sheeley}, Jr., N.~R.},
        title = "{Magnetic Flux Transport and the Sun's Dipole Moment: New Twists to the Babcock-Leighton Model}",
      journal = {\apj},
         year = 1991,
        month = jul,
       volume = {375},
        pages = {761},
          doi = {10.1086/170240},
       adsurl = {https://ui.adsabs.harvard.edu/abs/1991ApJ...375..761W}
}

@ARTICLE{BaumannAA2004,
       author = {{Baumann}, I. and {Schmitt}, D. and {Sch{\"u}ssler}, M. and {Solanki}, S.~K.},
        title = "{Evolution of the large-scale magnetic field on the solar surface: A parameter study}",
      journal = {\aap},
         year = 2004,
        month = nov,
       volume = {426},
        pages = {1075-1091},
          doi = {10.1051/0004-6361:20048024},
       adsurl = {https://ui.adsabs.harvard.edu/abs/2004A&A...426.1075B}
}

@ARTICLE{WhitbreadAA2017,
       author = {{Whitbread}, T. and {Yeates}, A.~R. and {Mu{\~n}oz-Jaramillo}, A. and {Petrie}, G.~J.~D.},
        title = "{Parameter optimization for surface flux transport models}",
      journal = {\aap},
         year = 2017,
        month = nov,
       volume = {607},
          eid = {A76},
        pages = {A76},
          doi = {10.1051/0004-6361/201730689},
archivePrefix = {arXiv},
       eprint = {1708.01098},
 primaryClass = {astro-ph.SR},
       adsurl = {https://ui.adsabs.harvard.edu/abs/2017A&A...607A..76W}
}

@ARTICLE{YeatesSSRv2023,
       author = {{Yeates}, Anthony R. and {Cheung}, Mark C.~M. and {Jiang}, Jie and {Petrovay}, Kristof and {Wang}, Yi-Ming},
        title = "{Surface Flux Transport on the Sun}",
      journal = {\ssr},
         year = 2023,
        month = jun,
       volume = {219},
       number = {4},
          eid = {31},
        pages = {31},
          doi = {10.1007/s11214-023-00978-8},
archivePrefix = {arXiv},
       eprint = {2303.01209},
 primaryClass = {astro-ph.SR},
       adsurl = {https://ui.adsabs.harvard.edu/abs/2023SSRv..219...31Y}
}

@ARTICLE{JiangAA2011II,
       author = {{Jiang}, J. and {Cameron}, R.~H. and {Schmitt}, D. and {Sch{\"u}ssler}, M.},
        title = "{The solar magnetic field since 1700. II. Physical reconstruction of total, polar and open flux}",
      journal = {\aap},
         year = 2011,
        month = apr,
       volume = {528},
          eid = {A83},
        pages = {A83},
          doi = {10.1051/0004-6361/201016168},
archivePrefix = {arXiv},
       eprint = {1102.1270},
 primaryClass = {astro-ph.SR},
       adsurl = {https://ui.adsabs.harvard.edu/abs/2011A&A...528A..83J}
}

@ARTICLE{YeatesSoPh2015,
       author = {{Yeates}, A.~R. and {Baker}, D. and {van Driel-Gesztelyi}, L.},
        title = "{Source of a Prominent Poleward Surge During Solar Cycle 24}",
      journal = {\solphys},
         year = 2015,
        month = nov,
       volume = {290},
       number = {11},
        pages = {3189-3201},
          doi = {10.1007/s11207-015-0660-9},
archivePrefix = {arXiv},
       eprint = {1502.04854},
 primaryClass = {astro-ph.SR},
       adsurl = {https://ui.adsabs.harvard.edu/abs/2015SoPh..290.3189Y}
}

@ARTICLE{VirtanenAA2017,
       author = {{Virtanen}, I.~O.~I. and {Virtanen}, I.~I. and {Pevtsov}, A.~A. and {Yeates}, A. and {Mursula}, K.},
        title = "{Reconstructing solar magnetic fields from historical observations. II. Testing the surface flux transport model}",
      journal = {\aap},
         year = 2017,
        month = jul,
       volume = {604},
          eid = {A8},
        pages = {A8},
          doi = {10.1051/0004-6361/201730415},
       adsurl = {https://ui.adsabs.harvard.edu/abs/2017A&A...604A...8V}
}

@ARTICLE{WangAA2021,
       author = {{Wang}, Zi-Fan and {Jiang}, Jie and {Wang}, Jing-Xiu},
        title = "{Algebraic quantification of an active region contribution to the solar cycle}",
      journal = {\aap},
         year = 2021,
        month = jun,
       volume = {650},
          eid = {A87},
        pages = {A87},
          doi = {10.1051/0004-6361/202140407},
archivePrefix = {arXiv},
       eprint = {2104.04307},
 primaryClass = {astro-ph.SR},
       adsurl = {https://ui.adsabs.harvard.edu/abs/2021A&A...650A..87W}
}

@ARTICLE{PalApJ2026,
       author = {{Pal}, Subhadip and {Hazra}, Gopal and {Mandal}, Sudip},
        title = "{Improved Reconstruction of the Century-long Solar Magnetic Field by Incorporating Morphological Asymmetry in Sunspots}",
      journal = {\apj},
         year = 2026,
        month = may,
       volume = {1002},
       number = {2},
          eid = {140},
        pages = {140},
          doi = {10.3847/1538-4357/ae5930},
archivePrefix = {arXiv},
       eprint = {2509.23691},
 primaryClass = {astro-ph.SR},
       adsurl = {https://ui.adsabs.harvard.edu/abs/2026ApJ..1002..140P}
}

@ARTICLE{UptonGeoRL2018,
       author = {{Upton}, Lisa A. and {Hathaway}, David H.},
        title = "{An Updated Solar Cycle 25 Prediction With AFT: The Modern Minimum}",
      journal = {\grl},
         year = 2018,
        month = aug,
       volume = {45},
       number = {16},
        pages = {8091-8095},
          doi = {10.1029/2018GL078387},
archivePrefix = {arXiv},
       eprint = {1808.04868},
 primaryClass = {astro-ph.SR},
       adsurl = {https://ui.adsabs.harvard.edu/abs/2018GeoRL..45.8091U}
}

@ARTICLE{CameronAA2012,
       author = {{Cameron}, R.~H. and {Schmitt}, D. and {Jiang}, J. and {I{\textcommabelow s}{\i}k}, E.},
        title = "{Surface flux evolution constraints for flux transport dynamos}",
      journal = {\aap},
         year = 2012,
        month = jun,
       volume = {542},
          eid = {A127},
        pages = {A127},
          doi = {10.1051/0004-6361/201218906},
archivePrefix = {arXiv},
       eprint = {1205.1136},
 primaryClass = {astro-ph.SR},
       adsurl = {https://ui.adsabs.harvard.edu/abs/2012A&A...542A.127C}
}

@ARTICLE{LuoAA2026,
       author = {{Luo}, Yukun and {Jiang}, Jie and {Li}, Binghang and {Zhang}, Zebin and {Wang}, Ruihui},
        title = "{Constraining the outer boundary condition for the Babcock-Leighton dynamo models}",
      journal = {\aap},
         year = 2026,
        month = jan,
       volume = {705},
          eid = {A237},
        pages = {A237},
          doi = {10.1051/0004-6361/202558038},
archivePrefix = {arXiv},
       eprint = {2512.09371},
 primaryClass = {astro-ph.SR},
       adsurl = {https://ui.adsabs.harvard.edu/abs/2026A&A...705A.237L}
}

@ARTICLE{BaumannAA2006,
       author = {{Baumann}, I. and {Schmitt}, D. and {Sch{\"u}ssler}, M.},
        title = "{A necessary extension of the surface flux transport model}",
      journal = {\aap},
         year = 2006,
        month = jan,
       volume = {446},
       number = {1},
        pages = {307-314},
          doi = {10.1051/0004-6361:20053488},
       adsurl = {https://ui.adsabs.harvard.edu/abs/2006A&A...446..307B}
}

@ARTICLE{PetrovayAA2019,
       author = {{Petrovay}, K. and {Talafha}, M.},
        title = "{Optimization of surface flux transport models for the solar polar magnetic field}",
      journal = {\aap},
         year = 2019,
        month = dec,
       volume = {632},
          eid = {A87},
        pages = {A87},
          doi = {10.1051/0004-6361/201936099},
archivePrefix = {arXiv},
       eprint = {1909.06125},
 primaryClass = {astro-ph.SR},
       adsurl = {https://ui.adsabs.harvard.edu/abs/2019A&A...632A..87P}
}

@ARTICLE{Chatterjee2026ApJ,
       author = {{Chatterjee}, Soumyadeep and {Hazra}, Gopal},
        title = "{Probing the Large-scale Magnetic Field inside the Sun from Three Decades of Observed Surface Magnetograms}",
      journal = {\apjl},
         year = 2026,
        month = jan,
       volume = {997},
       number = {1},
          eid = {L27},
        pages = {L27},
          doi = {10.3847/2041-8213/ae3138},
archivePrefix = {arXiv},
       eprint = {2509.23959},
 primaryClass = {astro-ph.SR},
       adsurl = {https://ui.adsabs.harvard.edu/abs/2026ApJ...997L..27C}
}

@ARTICLE{Schrijver2002ApJ,
       author = {{Schrijver}, Carolus J. and {De Rosa}, Marc L. and {Title}, Alan M.},
        title = "{What Is Missing from Our Understanding of Long-Term Solar and Heliospheric Activity?}",
      journal = {\apj},
         year = 2002,
        month = oct,
       volume = {577},
       number = {2},
        pages = {1006-1012},
          doi = {10.1086/342247},
       adsurl = {https://ui.adsabs.harvard.edu/abs/2002ApJ...577.1006S}
}

@ARTICLE{Elsasser1946PhRv,
       author = {{Elsasser}, Walter M.},
        title = "{Induction Effects in Terrestrial Magnetism Part I. Theory}",
      journal = {Physical Review},
         year = 1946,
        month = feb,
       volume = {69},
       number = {3-4},
        pages = {106-116},
          doi = {10.1103/PhysRev.69.106},
       adsurl = {https://ui.adsabs.harvard.edu/abs/1946PhRv...69..106E}
}

@INPROCEEDINGS{Hathway_2003ESAS,
       author = {{Hathaway}, David H.},
        title = "{Large scale flows through the solar cycle}",
    booktitle = {GONG+ 2002. Local and Global Helioseismology: the Present and Future},
         year = 2003,
       editor = {{Sawaya-Lacoste}, Huguette},
       series = {ESA Special Publication},
       volume = {517},
        month = feb,
        pages = {87-96},
       adsurl = {https://ui.adsabs.harvard.edu/abs/2003ESASP.517...87H}
}

@ARTICLE{Cameron2016,
       author = {{Cameron}, R.~H. and {Jiang}, J. and {Sch{\"u}ssler}, M.},
        title = "{Solar Cycle 25: Another Moderate Cycle?}",
      journal = {\apjl},
         year = 2016,
        month = jun,
       volume = {823},
       number = {2},
          eid = {L22},
        pages = {L22},
          doi = {10.3847/2041-8205/823/2/L22},
archivePrefix = {arXiv},
       eprint = {1604.05405},
 primaryClass = {astro-ph.SR},
       adsurl = {https://ui.adsabs.harvard.edu/abs/2016ApJ...823L..22C}
}

@ARTICLE{Jiang2018ApJ,
       author = {{Jiang}, Jie and {Wang}, Jing-Xiu and {Jiao}, Qi-Rong and {Cao}, Jin-Bin},
        title = "{Predictability of the Solar Cycle Over One Cycle}",
      journal = {\apj},
         year = 2018,
        month = aug,
       volume = {863},
       number = {2},
          eid = {159},
        pages = {159},
          doi = {10.3847/1538-4357/aad197},
archivePrefix = {arXiv},
       eprint = {1807.01543},
 primaryClass = {astro-ph.SR},
       adsurl = {https://ui.adsabs.harvard.edu/abs/2018ApJ...863..159J}
}

@ARTICLE{Wang2002ApJ,
       author = {{Wang}, Y.-M. and {Sheeley}, Jr., N.~R. and {Lean}, J.},
        title = "{Meridional Flow and the Solar Cycle Variation of the Sun's Open Magnetic Flux}",
      journal = {\apj},
         year = 2002,
        month = dec,
       volume = {580},
       number = {2},
        pages = {1188-1196},
          doi = {10.1086/343845},
       adsurl = {https://ui.adsabs.harvard.edu/abs/2002ApJ...580.1188W}
}

@ARTICLE{Snodgrass1990ApJ,
       author = {{Snodgrass}, Herschel B. and {Ulrich}, Roger K.},
        title = "{Rotation of Doppler Features in the Solar Photosphere}",
      journal = {\apj},
         year = 1990,
        month = mar,
       volume = {351},
        pages = {309},
          doi = {10.1086/168467},
       adsurl = {https://ui.adsabs.harvard.edu/abs/1990ApJ...351..309S}
}

@ARTICLE{Schssler2006AAP,
       author = {{Sch{\"u}ssler}, M. and {Baumann}, I.},
        title = "{Modeling the Sun's open magnetic flux}",
      journal = {\aap},
         year = 2006,
        month = dec,
       volume = {459},
       number = {3},
        pages = {945-953},
          doi = {10.1051/0004-6361:20065871},
       adsurl = {https://ui.adsabs.harvard.edu/abs/2006A&A...459..945S}
}

@misc{sftyeates,
  author       = {{Anthony Yeates}},
  year         = {2016},
  title        = {sft\_data},
  howpublished = {\url{https://github.com/antyeates1983/sft_data}},
  note         = {[accessed: July 2026]}
}

@ARTICLE{HanasogeLRSP2022,
       author = {{Hanasoge}, Shravan M.},
        title = "{Surface and interior meridional circulation in the Sun}",
      journal = {Living Reviews in Solar Physics},
         year = 2022,
        month = dec,
       volume = {19},
       number = {1},
          eid = {3},
        pages = {3},
          doi = {10.1007/s41116-022-00034-7},
       adsurl = {https://ui.adsabs.harvard.edu/abs/2022LRSP...19....3H}
}

@ARTICLE{GolubevaMNRAS2023,
       author = {{Golubeva}, Elena M. and {Biswas}, Akash and {Khlystova}, Anna I. and {Kumar}, Pawan and {Karak}, Bidya Binay},
        title = "{Probing the variations in the timing of the Sun's polar magnetic field reversals through observations and surface flux transport simulations}",
      journal = {\mnras},
         year = 2023,
        month = oct,
       volume = {525},
       number = {2},
        pages = {1758-1768},
          doi = {10.1093/mnras/stad2254},
archivePrefix = {arXiv},
       eprint = {2307.13452},
 primaryClass = {astro-ph.SR},
       adsurl = {https://ui.adsabs.harvard.edu/abs/2023MNRAS.525.1758G}
}

@ARTICLE{PalApJ2023,
       author = {{Pal}, Shaonwita and {Bhowmik}, Prantika and {Mahajan}, Sushant S. and {Nandy}, Dibyendu},
        title = "{Impact of Anomalous Active Regions on the Large-scale Magnetic Field of the Sun}",
      journal = {\apj},
         year = 2023,
        month = aug,
       volume = {953},
       number = {1},
          eid = {51},
        pages = {51},
          doi = {10.3847/1538-4357/acd77e},
archivePrefix = {arXiv},
       eprint = {2305.13145},
 primaryClass = {astro-ph.SR},
       adsurl = {https://ui.adsabs.harvard.edu/abs/2023ApJ...953...51P}
}

@book{Brent1973,
  author    = {{Brent}, R.~P.},
  title     = {{Algorithms for Minimization without Derivatives}},
  publisher = {NJ: Prentice-Hall},
  address   = {Englewood Cliffs, NJ},
  year      = {1973},
  isbn      = {0-13-022335-2}
}

@ARTICLE{TemmerLRSP2021,
       author = {{Temmer}, Manuela},
        title = "{Space weather: the solar perspective: An update to Schwenn (2006)}",
      journal = {Living Reviews in Solar Physics},
         year = 2021,
        month = dec,
       volume = {18},
       number = {1},
          eid = {4},
        pages = {4},
          doi = {10.1007/s41116-021-00030-3},
archivePrefix = {arXiv},
       eprint = {2104.04261},
 primaryClass = {astro-ph.SR},
       adsurl = {https://ui.adsabs.harvard.edu/abs/2021LRSP...18....4T}
}

@ARTICLE{Chou11,
       author = {{Choudhuri}, Arnab Rai},
        title = "{The origin of the solar magnetic cycle}",
      journal = {Pramana},
         year = 2011,
        month = jul,
       volume = {77},
       number = {1},
        pages = {77-96},
          doi = {10.1007/s12043-011-0113-4},
archivePrefix = {arXiv},
       eprint = {1103.3385},
 primaryClass = {astro-ph.SR},
       adsurl = {https://ui.adsabs.harvard.edu/abs/2011Prama..77...77C}
}

@ARTICLE{HCM17,
       author = {{Hazra}, Gopal and {Choudhuri}, Arnab Rai and {Miesch}, Mark S.},
        title = "{A Theoretical Study of the Build-up of the Sun{\textquoteright}s Polar Magnetic Field by using a 3D Kinematic Dynamo Model}",
      journal = {The Astrophysical Journal},
         year = 2017,
        month = jan,
       volume = {835},
       number = {1},
          eid = {39},
        pages = {39},
          doi = {10.3847/1538-4357/835/1/39},
archivePrefix = {arXiv},
       eprint = {1610.02726},
 primaryClass = {astro-ph.SR},
       adsurl = {https://ui.adsabs.harvard.edu/abs/2017ApJ...835...39H}
}

@ARTICLE{HM18,
       author = {{Hazra}, Gopal and {Miesch}, Mark S.},
        title = "{Incorporating Surface Convection into a 3D Babcock-Leighton Solar Dynamo Model}",
      journal = {The Astrophysical Journal},
         year = 2018,
        month = sep,
       volume = {864},
       number = {2},
          eid = {110},
        pages = {110},
          doi = {10.3847/1538-4357/aad556},
archivePrefix = {arXiv},
       eprint = {1804.03100},
 primaryClass = {astro-ph.SR},
       adsurl = {https://ui.adsabs.harvard.edu/abs/2018ApJ...864..110H}
}

@ARTICLE{Iijima2019,
       author = {{Iijima}, H. and {Hotta}, H. and {Imada}, S.},
        title = "{Effect of Morphological Asymmetry between Leading and Following Sunspots on the Prediction of Solar Cycle Activity}",
      journal = {The Astrophysical Journal},
         year = 2019,
        month = sep,
       volume = {883},
       number = {1},
          eid = {24},
        pages = {24},
          doi = {10.3847/1538-4357/ab3b04},
archivePrefix = {arXiv},
       eprint = {1908.04474},
 primaryClass = {astro-ph.SR},
       adsurl = {https://ui.adsabs.harvard.edu/abs/2019ApJ...883...24I}
}

@ARTICLE{KM17,
       author = {{Karak}, Bidya Binay and {Miesch}, Mark},
        title = "{Solar Cycle Variability Induced by Tilt Angle Scatter in a Babcock-Leighton Solar Dynamo Model}",
      journal = {The Astrophysical Journal},
         year = 2017,
        month = sep,
       volume = {847},
       number = {1},
          eid = {69},
        pages = {69},
          doi = {10.3847/1538-4357/aa8636},
archivePrefix = {arXiv},
       eprint = {1706.08933},
 primaryClass = {astro-ph.SR},
       adsurl = {https://ui.adsabs.harvard.edu/abs/2017ApJ...847...69K}
}

@ARTICLE{Pulkkinen2007,
       author = {{Pulkkinen}, Tuija},
        title = "{Space Weather: Terrestrial Perspective}",
      journal = {Living Reviews in Solar Physics},
         year = 2007,
        month = dec,
       volume = {4},
       number = {1},
          eid = {1},
        pages = {1},
          doi = {10.12942/lrsp-2007-1},
       adsurl = {https://ui.adsabs.harvard.edu/abs/2007LRSP....4....1P}
}

@ARTICLE{Schwenn2006,
       author = {{Schwenn}, Rainer},
        title = "{Space Weather: The Solar Perspective}",
      journal = {Living Reviews in Solar Physics},
         year = 2006,
        month = dec,
       volume = {3},
       number = {1},
          eid = {2},
        pages = {2},
          doi = {10.12942/lrsp-2006-2},
       adsurl = {https://ui.adsabs.harvard.edu/abs/2006LRSP....3....2S}
}

@ARTICLE{Yeates2020,
       author = {{Yeates}, Anthony R.},
        title = "{How Good Is the Bipolar Approximation of Active Regions for Surface Flux Transport?}",
      journal = {\solphys},
         year = 2020,
        month = sep,
       volume = {295},
       number = {9},
          eid = {119},
        pages = {119},
          doi = {10.1007/s11207-020-01688-y},
archivePrefix = {arXiv},
       eprint = {2008.03203},
 primaryClass = {astro-ph.SR},
       adsurl = {https://ui.adsabs.harvard.edu/abs/2020SoPh..295..119Y}
}

@ARTICLE{Charbonneau2020,
       author = {{Charbonneau}, Paul},
        title = "{Dynamo models of the solar cycle}",
      journal = {Living Reviews in Solar Physics},
         year = 2020,
        month = jun,
       volume = {17},
       number = {1},
          eid = {4},
        pages = {4},
          doi = {10.1007/s41116-020-00025-6},
       adsurl = {https://ui.adsabs.harvard.edu/abs/2020LRSP...17....4C}
}

@ARTICLE{yeates2008,
       author = {{Yeates}, Anthony R. and {Nandy}, Dibyendu and {Mackay}, Duncan H.},
        title = "{Exploring the Physical Basis of Solar Cycle Predictions: Flux Transport Dynamics and Persistence of Memory in Advection- versus Diffusion-dominated Solar Convection Zones}",
      journal = {The Astrophysical Journal},
         year = 2008,
        month = jan,
       volume = {673},
       number = {1},
        pages = {544-556},
          doi = {10.1086/524352},
archivePrefix = {arXiv},
       eprint = {0709.1046},
 primaryClass = {astro-ph},
       adsurl = {https://ui.adsabs.harvard.edu/abs/2008ApJ...673..544Y}
}

@ARTICLE{schatten1978,
       author = {{Schatten}, K.~H. and {Scherrer}, P.~H. and {Svalgaard}, L. and {Wilcox}, J.~M.},
        title = "{Using Dynamo Theory to predict the sunspot number during Solar Cycle 21}",
      journal = {\grl},
         year = 1978,
        month = may,
       volume = {5},
       number = {5},
        pages = {411-414},
          doi = {10.1029/GL005i005p00411},
       adsurl = {https://ui.adsabs.harvard.edu/abs/1978GeoRL...5..411S}
}

@ARTICLE{petrovay2020,
       author = {{Petrovay}, Krist{\'o}f},
        title = "{Solar cycle prediction}",
      journal = {Living Reviews in Solar Physics},
         year = 2020,
        month = mar,
       volume = {17},
       number = {1},
          eid = {2},
        pages = {2},
          doi = {10.1007/s41116-020-0022-z},
archivePrefix = {arXiv},
       eprint = {1907.02107},
 primaryClass = {astro-ph.SR},
       adsurl = {https://ui.adsabs.harvard.edu/abs/2020LRSP...17....2P}
}

@ARTICLE{bhowmik2018,
   author = {{Bhowmik}, P. and {Nandy}, D.},
    title = "{Prediction of the strength and timing of sunspot cycle 25 reveal decadal-scale space environmental conditions}",
  journal = {Nature Communications},
     year = 2018,
    month = dec,
   volume = 9,
    pages = {5209},
      doi = {10.1038/s41467-018-07690-0}
}

@ARTICLE{nandy2021b,
       author = {{Nandy}, Dibyendu and {Martens}, Petrus C.~H. and {Obridko}, Vladimir and {Dash}, Soumyaranjan and {Georgieva}, Katya},
        title = "{Solar evolution and extrema: current state of understanding of long-term solar variability and its planetary impacts}",
      journal = {Progress in Earth and Planetary Science},
         year = 2021,
        month = dec,
       volume = {8},
       number = {1},
          eid = {40},
        pages = {40},
          doi = {10.1186/s40645-021-00430-x},
       adsurl = {https://ui.adsabs.harvard.edu/abs/2021PEPS....8...40N}
}


\end{document}